\documentclass[colorlinks=true, linkcolor=blue, citecolor=blue, urlcolor=blue]{aa}
\usepackage{enumitem}
\usepackage[switch]{lineno}
\usepackage[utf8]{inputenc}
\usepackage{graphicx}
\usepackage{multirow}
\usepackage{pifont}
\usepackage{txfonts}
\usepackage{natbib}
\usepackage{footnote}
\usepackage{longtable}
\usepackage{multirow}
\usepackage{booktabs}
\usepackage{xcolor}
\usepackage{changepage}
\usepackage{amssymb}
\usepackage{amsmath}
\usepackage{hyperref}
\usepackage{amsfonts}
\usepackage{geometry}
\usepackage{longtable}
\usepackage{multirow}
\usepackage{booktabs}
\usepackage{pifont}
\usepackage{tablefootnote}
\usepackage{float}
\usepackage{xcolor}
\begin{document}

\title{From X-rays to High-Energy Gamma-rays: A Comprehensive Multi-Wavelength Study of Early Gamma-Ray Burst Afterglows}
\titlerunning{MWL Study of GRB Afterglows}

\author{Pawan Tiwari\thanks{\href{mailto:pawan.tiwari@gssi.it}{pawan.tiwari@gssi.it}} \inst{1,2},
Biswajit Banerjee \inst{1,2,3}, 
Davide Miceli\inst{4,5}, 
Gor Oganesyan\inst{1,2,3},
Annarita Ierardi\inst{1,2}, 
Samanta Macera\inst{1,2},  
Marica Branchesi\inst{1,2,3}, 
Lara Nava\inst{5,6}, 
Shraddha Mohnani\inst{7},
Sushmita Agarwal\inst{7}, and
Amit Shukla\inst{7} }

\institute{
\inst{1} Gran Sasso Science Institute (GSSI), Viale F. Crispi 7, L’Aquila (AQ), I-67100, Italy\\
\inst{2} INFN - Laboratori Nazionali del Gran Sasso, L’Aquila (AQ), I-67100, Italy\\
\inst{3} INAF - Osservatorio Astronomico d’Abruzzo, Via M. Maggini snc, I-64100 Teramo, Italy \\
\inst{4} Istituto Nazionale di Fisica Nucleare (INFN), Sez. Padova, Via Marzolo 8, 35131 Padova, Italy \\
\inst{5} INAF - Osservatorio Astronomico di Brera, Via Emilio Bianchi 46, 23807 Merate, Italy \\
\inst{6} Istituto Nazionale di Fisica Nucleare (INFN), Sez. Trieste, Via A. Valerio 2, 34100 Trieste, Italy \\
\inst{7} Department of Astronomy, Astrophysics and Space Engineering, Indian Institute of Technology Indore, India \\
}

\authorrunning{P. Tiwari et al.}   

\abstract
{Gamma-ray Bursts (GRBs) generate powerful relativistic jets that inject a large amount of energy into their surrounding environment, producing blast waves that accelerate particles to high energies. The GRB afterglow radiation provides a powerful means to investigate the microphysics of relativistic shocks and to probe the medium surrounding the progenitor of the burst. In this study, we present a comprehensive multiwavelength analysis of 31 GRBs observed between 2008 and 2024 from the Neil Gehrels Swift Observatory (X-ray Telescope and Burst Alert Telescope) and the Fermi Large Area Telescope, covering photon energies from 0.3 keV to 300 GeV. Our goal is to characterize the broadband spectral properties of GRB afterglows in soft X-rays, hard X-rays, and high-energy gamma rays. We investigate correlations between spectral shape and energy output across different parts of the spectrum. The observed emission is modeled using a forward shock scenario that includes both synchrotron and synchrotron self-Compton (SSC) radiation losses. The results favor an SSC-dominated radiation model in a wind-like medium, consistent with expectations for long-duration GRBs. Crucially, this work provides new benchmarks for the microphysical parameters governing the emission, particularly indicating a notably low magnetic energy fraction, which refines previous estimates.
 By modeling broadband data, this study offers one of the most detailed SSC analyses in a wind-like environment to date.
Notably, our results naturally account for the comparable energy output observed in both the soft X-ray and TeV bands, consistent with the previously detected TeV-GRBs.}

\keywords{high energy astrophysics, gamma rays: bursts, gamma rays: observations, methods: observational}

\maketitle

\section{Introduction}
Gamma-ray bursts (GRBs) are one of the most energetic and luminous transient phenomena in the Universe. These events are observed from cosmological distances and characterized by an initial highly variable prompt emission phase followed by a multi-wavelength afterglow. The prompt emission typically lasts from a few sub-seconds to several minutes and predominantly occurs in the keV to MeV energy range. Subsequently, GRBs exhibit long-lasting afterglow emission that can span from minutes to months and is observed across the entire electromagnetic spectrum, from radio to very-high-energy (VHE; E $>$100 GeV) gamma-rays. 

GRB afterglows provide critical insights into the kinetic energy remaining in the jet after the prompt emission phase, the profile of the circumburst medium~\citep{2000ApJ...536..195C}, and the microphysical processes governing shock dynamics~\citep{1997ApJ...476..232M, 1998ApJ...497L..17S}. Multiwavelength observations, particularly in the X-ray to MeV gamma-ray bands, are crucial for constraining the characteristic synchrotron frequencies~\citep{1998ApJ...497L..17S}. High-energy (HE; E$>$100 MeV) gamma-ray photons are thought to originate from the most energetic electrons, which cool rapidly on the dynamical timescale of the shock~\citep{2010MNRAS.409..226K, 2010MNRAS.403..926G}. Importantly, GeV afterglow emission has been shown to be essential to infer the kinetic energy of the jet post-prompt emission phase~\citep{2015MNRAS.454.1073B}. Thanks to the Fermi Large Area Telescope (LAT; 30\,MeV to $>$300\,GeV), such HE afterglows can now be routinely observed~\citep{Atwood2009}. LAT-detected GRB afterglows spectra often display a long-lasting power-law decay \citep{2010MNRAS.403..926G, Fermi-LAT:2013oro}, which is thought to be originated from synchrotron emission from external forward shocks~\citep{2010MNRAS.409..226K, 2010MNRAS.403..926G}. However, theoretical models have long predicted the emergence of a synchrotron self-Compton (SSC) component at HE~\citep{Dermer:1999eh, 2001ApJ...548..787S, 2001ApJ...559..110Z}. Moreover, independent studies of GRB afterglows in the keV~\citep{2012MNRAS.425..506D} and GeV~\citep{2014MNRAS.443.3578N} energies have revealed a consistent temporal decay of the ratio between isotropic luminosity ($L_{\rm ISO}$) and isotropic energy ($E^{\rm prompt}_{\rm ISO}$) with time, suggesting a common underlying radiative mechanism across energy bands.

In 2019, observations with the Major Atmospheric Gamma-ray Imaging Cherenkov (MAGIC) telescope of GRB190114C~\citep{MAGIC:2019irs}, and with H.E.S.S. of GRB180720B~\citep{2019Natur.575..464A}, showed that GRBs can emit photons up to TeV energies. The periods of TeV detections for GRB190114C were accompanied by simultaneous observations of the keV and GeV data, demonstrating the presence of a double spectral component, which was interpreted as SSC~\citep{MAGIC:2019lau}. These simultaneous multi-wavelength observations from X-ray to VHE emphasize the importance of broadband modeling. In 2022, the Large High Altitude Air Shower Observatory (LHAASO) reported the detection of GRB221009A, known as the "Brightest of All Time" ~\citep[BOAT;][]{2023Sci...380.1390L}. This exceptional burst showed clear signs of two spectral components in the afterglow~\citep{Banerjee:2024hxp, 2024MNRAS.529L..47K} thanks to the simultaneous detection of the MeV and GeV afterglow. Moreover, the keV X-ray band serves as a probe of the underlying synchrotron emission, while the GeV band provides direct evidence for SSC radiation in the afterglow phase.
In addition, \citet{Macera:2025wrv} analyzed the prompt emission spectrum of a sample of GRBs showing high-energy radiation during the prompt emission phase, and found that some exhibit a hint of a second component in the 100 MeV - 10 GeV energy range during the early phases, which might correspond to an early afterglow SSC emission, though its peak remained unconstrained.

Although only five GRBs, including GRB 190829A~\citep{2021Sci...372.1081H} and GRB 201216C~\citep{2024MNRAS.527.5856A} have been firmly detected in the VHE domain, these events have significantly advanced our understanding of broadband afterglow emission. In contrast, Fermi/LAT has detected more than 250 GRBs in the GeV range over the past 16 years, thanks to its wide field of view and continuous all-sky monitoring~\citep{2009ApJ...697.1071A}. Similarly, the Swift X-ray Telescope (XRT, 0.3–10~keV) aboard the Neil Gehrels Swift Observatory has detected over 2,100 GRBs since the beginning of its operations. Together, XRT and LAT span seven orders of magnitude in energy, and the joint detections provide an unparalleled opportunity for comprehensive multi-wavelength studies. Several joint X-ray and GeV studies \citep{2015MNRAS.454.1073B, 2018ApJ...863..138A} have significantly advanced our understanding of the afterglow phase. Their results show that the broadband data can be explained by either a synchrotron-only model or an SSC scenario, with both interpretations remaining compatible with the observed spectra. Moreover, the previous studies were carried out before the discoveries of the afterglow emission without TeV detection, which further supports the presence of SSC. The studies reveals no distinct preference between a uniform interstellar medium (ISM) or a wind-like environment. Moreover, both works were limited by incomplete broadband spectral coverage, especially at hard X-ray ($>$ 10\,keV).

Differently from previous studies, we performed a systematic time-resolved spectral analysis of multiwavelength early afterglow (up to 10,000 seconds after the burst) including X-ray data from the XRT and GeV data from LAT between August 2008 and 2024. By combining data from the X-ray and GeV energy ranges, we compare the simultaneous energy-fluxes \footnote{In the rest of the paper, we denote the energy-flux by flux, when not mentioned otherwise.} (erg cm$^{-2}$ s$^{-1}$) and spectral indices. We also included hard X-ray data from the Swift Burst Alert Telescope (BAT; 15–150~keV) when available in order to expand the spectra up to 150\,keV and constrain potential spectral peaks. 

In this work, we employ the SSC model based on the Leptonic Modeling Code~\citep[LeMoC;][]{Stathopoulos:2023qoy} to explain the emission mechanism responsible for producing X-ray and HE gamma-ray photons. Moreover, we investigate whether a single set of microphysical parameters can account for the afterglow trends observed in both HE gamma rays \citep{2014MNRAS.443.3578N} and X-rays \citep{2012MNRAS.425..506D}, aiming to enforce a unified interpretation across different energy bands. This investigation also considers the impact of the circumburst environment, specifically comparing constant-density ISM and wind-like density profiles. Additionally, we test the hypothesis that a preferred set of parameters can reproduce the observed flux and spectral index correlations. Finally, we predict the VHE emission and examine its correlation with lower-energy counterparts. Although afterglow parameters can vary significantly from one burst to another, studying a population allows us to assume the existence of benchmark parameter values that broadly characterize the shock dynamics and emission properties.

This paper is structured as follows: in Sect.~\ref{sec:sample_definition}, we describe the sample of GRBs selected for this study, together with the selection criteria. Section~\ref{sect:mwl_analysis} outlines the methodology adopted for the analysis of data from XRT, BAT, and LAT. Section~\ref{sec:Time-bin selection} explains the strategy used to select time intervals for each GRB, based on the observation modes of the respective instruments. The results obtained from our multi-wavelength analysis are presented in Sect.~\ref{sec:results}. An interpretation of these findings is provided in Sect.~\ref{sec:Intrepretation}. Finally, Sect.~\ref{Discussion} summarizes our main results and offers a discussion of their possible implications.

\section{Sample Definition} \label{sec:sample_definition}
With 16 years of data from August 2008--August 2024, we selected a sample of GRBs that have simultaneous observations with XRT (0.3--10~keV), BAT (15--150~keV), and LAT (0.1--10~GeV). For each GRB, we require that these observations occur after \( T_{95}^\text{BAT} \) (\( T_{95}^\text{BAT} \) denotes the end time of the interval containing 90\% of the burst's total counts in the energy band 15--350~keV) and within 10\,ks seconds after the burst. In addition to this, and independently of the X-ray observations, we identified all GRBs with LAT detection significance greater than $4\sigma$ within 10\,ks of the burst. This resulted in a total of 257 GRBs, which are divided into two subsets:
\begin{itemize}
    \item August 2008--August 2018: 186 GRBs from the LAT second catalog~\citep{2019ApJ...878...52A}.
    \item September 2018--August 2024: 71 GRBs with confirmed LAT detections that are triggered in BAT\footnote{\url{https://swift.gsfc.nasa.gov/results/batgrbcat/}}\textsuperscript{,}\footnote{These BAT-detected GRBs are selected to ensure early follow-up by XRT and hence the possibility of having more joint detections with LAT.}. These GRBs are identified through a dedicated analysis described in Sect.~\ref{sec:LATanalysis}, using the localization (R.A.: right ascension and Dec.: declination) and trigger times provided by BAT.
\end{itemize}
This selection criteria yields 31 GRBs with at least one time bin exhibiting simultaneous XRT, BAT, and LAT observations within 10\,ks. We excluded time bins containing flares identified by the automated Swift analysis pipeline \citep{Willingale:2006zh}. GRBs with identified flare intervals are cataloged in Tab.~\ref{tab:Flare_GRBs}. Further details on time-bin selection criteria based on LAT and XRT observation modes are provided in Sect.~\ref{sec:Time-bin selection}. The final sample of GRBs analyzed in this study is presented in Tab.~\ref{tab:Sample1}.

\begin{figure}[h]
    \centering
    \includegraphics[width=0.90\columnwidth]{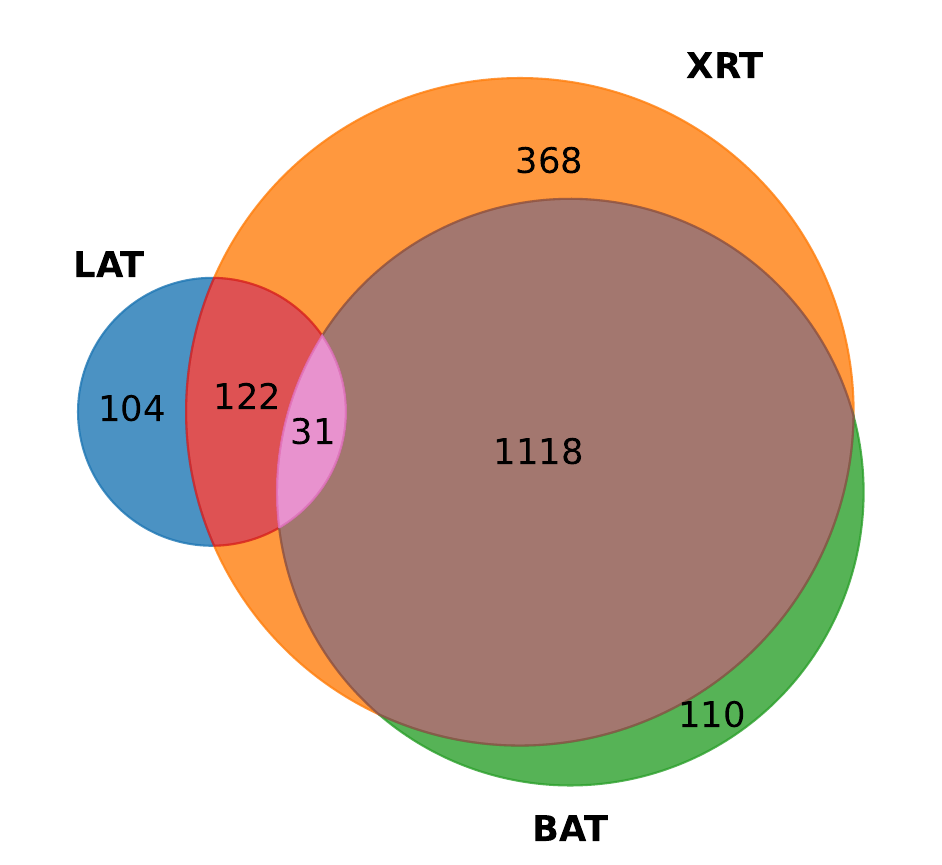}
    \caption{Venn diagram showing the number of GRBs detected by XRT, BAT, and LAT during the period from August 2008 to August 2024. The overlaps represent GRBs detected by multiple instruments. Specifically, 31 GRBs were jointly detected in at least one time interval by all three instruments, as evidenced in the pink overlap region.}
    \label{fig:GRB_detection}
\end{figure}

\begin{table*}[ht!]
\small
\centering
\renewcommand{\arraystretch}{1.2}
\setlength{\tabcolsep}{1.42pt} 
\begin{tabular}{|c|c|c|c|r|r|r|r|r|r|r|r|}
\hline
 \multirow{2}{*}{{GRB}} & \multirow{2}{*}{{Fermi}} & \multirow{3}{*}{{RA, Dec}} & \multirow{3}{*}{\text{z}} & \multirow{3}{*}{\( T_{0}^{\text{BAT}} \)} & \multirow{3}{*}{\( T_{0}^{\text{GBM}} \)} & \multirow{2}{*}{$\Delta {\rm T}_{0}$} &\multirow{2}{*}{ \( T_{90}^{\text{BAT}}\) }& \multirow{2}{*}{\( T_{90}^{\text{GBM}}\)} & \multirow{2}{*}{\( T_{95}^ \text{BAT}\)} & \multirow{2}{*}{\( T_{95}^ \text{GBM}\)} & $S_{\gamma}$  \\
    name & name  &  &  &  &  &     &     &      &     &      & (1-1000 keV) \\
         &       &  &  &  &  & [s] & [s] & [s]  & [s] & [s]  & $\times10^{-5}$[erg\,cm$^{-2}$] \\
    \hline
    081203A &           & 233.03, 63.50  & 2.05 & 13:57:11.6 & 13:57:11 & 0.6  &$223.0\pm89.9$ &$214.0\pm5.7$ &208.1 & 213.4 &     \\
    090510  & 090510016 & 333.55, -26.60 & 0.90 & 00:23:00.5 & 00:22:59 & 1.5  &$5.7\pm 1.9$   &$1.0\pm0.1$    &  5.7 & -0.6  & $0.34\pm 0.01$ \\
    091127A & 091127976 & 36.57, -18.95  & 0.49 & 23:25:45.8 & 23:25:45 & 0.8  &$7.0\pm0.2$    &$8.7\pm0.6$    &  6.8 & 7.9   & $2.07\pm0.01$\\
    100728A & 100728095 & 88.74, -15.26  & 1.57 & 02:18:24.2 & 02:17:30 &54.2  &$193.4\pm10.6$ &$165.4\pm2.9$  &153.7 & 124.5 & $12.70\pm0.10$\\
    110213A & 110213220 & 43.00, 49.29   & 1.46 & 05:17:29.5 & 05:17:11 &18.5  &$48.0\pm16.0$  &$33.5\pm1.6$   & 16.8& 15.8   & $0.94\pm0.01$\\
    110625A & 110625881 & 286.75, 6.75   &      & 21:08:28.4 & 21:08:18 &10.4  &$42.5\pm8.9$   &$30.7\pm0.6$   & 39.5& 16.5   & $6.55\pm0.01$\\
    110731A & 110731465 &280.52, -28.55  & 2.83 & 11:09:30.5 & 11:09:29 & 1.5  &$40.9\pm11.6$  &$7.5\pm0.6$    & 40.7 & 6.0   & $2.29\pm0.01$\\
    120729A & 120729456 & 13.05, 49.94   & 0.8  & 10:56:14.1 & 10:56:12 & 2.1  &$93.9\pm36.6$  &$24.5\pm2.6$   & 92.5& 23.4   & $0.51\pm0.01$\\
    121011A & 121011469 & 260.20, 1.14   &      & 11:15:30.4 & 11:15:25 & 5.4  &$97.2\pm17.2$  &$66.8\pm4.4$   & 92.7 & 60.4  & $0.40\pm0.04$\\
    130427A & 130427324 & 173.14, 27.69  & 0.34 & 07:47:57.5 & 07:47:06 &51.5  &$244.3\pm4.7$  &$138.2\pm3.2$  &197.3 & 90.8  & $246.00\pm1.00$\\
    140102A & 140102887 & 211.90, 1.33   &      & 21:17:37.8 & 21:17:37 & 0.8  &$55.1\pm15.4$  &$3.7\pm0.1$    & 55.5 & 3.3   & $1.78\pm0.01$\\
    140323A & 140323433 & 356.88, -79.91 &      & 10:23:11.9 & 10:22:53 &18.9  &$106.5\pm3.2$  &$116.5\pm3.01$ &102.3 & 92.5  & $3.24\pm0.02$\\
    150314A & 150314205 & 126.65, 63.83  & 1.76 & 04:54:52.9 & 04:54:50 & 2.9  &$14.8\pm2.6$   &$11.3\pm0.2$   & 13.6 & 7.8   & $8.15\pm0.01$ \\
    150403A & 150403913 & 311.50, -62.70 & 2.06 & 21:54:16.8 & 21:54:10 & 6.8  &$37.3\pm11.7$  &$22.3\pm0.8$   & 29.6 & 18.8  & $5.47\pm0.01$\\
    151006A & 151006413 & 147.45, 70.51  &      & 09:55:01.9 & 09:54:57 & 4.9  &$211.1\pm39.1$ &$95.0\pm6.1$   &209.4 & 88.5  & $1.23\pm0.01$ \\
    160325A & 160325291 & 15.60, -72.71  &      & 07:00:03.6 & 06:59:21 &42.6  &$61.7\pm12.1$  &$43.0\pm0.6$   & 20.8 & 2.4   & $1.86\pm0.02$\\
    160905A & 160905471 & 162.24, -50.80 &      & 11:18:58.4 & 11:18:55 & 3.4  &$64.0\pm16.0$  &$33.5\pm1.4$   & 48.8 & 34.0  & $7.32\pm0.04$ \\
    160917A & 160917479 & 295.67, 46.39  &      & 11:30:19.3 & 11:30:19 & 0.3  &$14.5\pm1.8$   &$19.2\pm19.5$  &  14.6 & 19.1  & $0.52\pm0.01$\\
    170405A & 170405777 & 219.81, -25.24 & 3.51 & 18:39:48.4 & 18:39:22 &26.4  &$165.3\pm32.9$ &$78.6\pm0.6$   &148.5 & 59.6  & $7.39\pm0.01$\\
    170728B & 170728961 & 238.04, 70.14  & 1.27 & 23:03:19.4 & 23:03:19 & 0.4  &$47.7\pm26.5$  &$46.3\pm0.8$   & 47.7 & 45.9  & $0.46\pm0.03$\\
    170813A & 170813051 & 201.05, -5.46  &      & 01:13:16.5 & 01:13:08 & 8.5  &$72.5\pm12.1$  &$111.4\pm4.9$  &65.1 & 103.4 & $0.36\pm0.01$\\
    170906A & 170906030 & 203.99, -47.12 &      & 00:43:11.6 & 00:43:08 & 3.6  &$88.1\pm1.4$   &$78.9\pm0.7$   & 96.1 & 87.3  & $9.47\pm0.01$\\
    171120A & 171120556 & 163.79, 22.45  &      & 13:20:02.5 & 13:20:02 & 0.5  &$64.0\pm16.0$  &$44.1\pm0.4$   & 48.5 & 43.6  & $1.61\pm0.01$\\
    180720B & 180720598 & 0.53, -2.93    & 0.65 & 14:21:44.6 & 14:21:39 & 5.6  &$108.4\pm3.7$  &$48.9\pm0.4$   &108.6 & 47.7  & $29.80\pm0.00$\\
    181020A & 181020792 & 13.96, -47.37  & 2.94 & 19:00:33.2 & 19:00:33 & 0.2  &$238.0\pm11.6$ &$15.1\pm0.6$   &238.8 & 14.4  & $2.81\pm0.00$\\
    190511A & 190511302 & 126.43, -20.24 &      & 07:14:48.3 & 07:14:24 & 24.0 &$27.7\pm2.4$   &$27.6\pm1.2$   &5.9 & 1.1  & $2.10\pm0.01$\\
    200716C & 200716957 & 196.01, 29.62  &      & 22:57:41.2 & 22:57:41 & 0.2  &$86.6\pm15.2$  &$5.3\pm0.4$    &86.5 & 5.0  & $0.96\pm0.00$\\
    210410A & 210410037 & 269.75, 45.37  &      & 00:53:16   & 00:53:16 & 0.0  &$52.9\pm4.2$   &$48.1\pm2.8$   &53.5 &   48.1& $3.10\pm0.01$\\
    210619B & 210619999 & 319.71, 33.85  & 1.94 & 23:59:25.1 & 23:59:25 & 0.1  &$60.9\pm0.3$   &$54.8\pm0.6$   &61.7 &   54.1& $30.25\pm0.01$\\
    220101A & 220101215 & 1.37, 31.75    & 4.62 & 05:10:11.7 & 05:10:12 & -0.2 &$161.9\pm12.7$ &$128.3\pm15.8$ &164.9 &  111.0& $6.04\pm0.02$\\
    240825A & 240825662 & 344.55, 1.03   & 0.66 & 15:52:59.8 & 15:53:00 & -0.2 &$416.0\pm48.0$ &$3.9\pm0.1$    &416.1 &  3.1& $10.02\pm0.00$\\
\hline
\end{tabular}
\caption{Sample of LAT-detected GRBs with simultaneous XRT observations within 10\,ks after the burst trigger in BAT (for period August 2008-2024). \(T_0^{\text{BAT}}\) is the trigger time in BAT, whereas \(T_{90}^{\text{BAT}}\) is the time interval between the detection of 5\% and 95\% of the total photons by BAT. Similarly \( T_{90}^{\text{GBM}} \) denotes for Fermi GBM. \(T_{95}^{\text{BAT}} \) and \(T_{95}^{\text{GBM}} \) denotes the end time of interval containing 90\% of burst's total count in BAT and GBM, respectively. In the table, \( T_{90}^{\text{BAT}} \), \(T_{95}^{\text{BAT}} \) , and \(T_{95}^{\text{GBM}} \) are reported relative to \( T_0^{\text{BAT}} \), while \( T_{90}^{\text{GBM}} \) is reported relative to \( T_0^{\text{GBM}}\)~\citep{2019ApJ...878...52A}. Fluence ($S_{\gamma}$; 1-1000 keV) is reported from ~\url{https://www.mpe.mpg.de/~jcg/grbgen.html}. The \(T_0^{\text{BAT}}\), \( T_{90}^{\text{BAT}} \), and \(T_{95}^{\text{BAT}} \) values are reported from ~\url{https://swift.gsfc.nasa.gov/results/batgrbcat/}.} Note that GRB 081203A was detected by BAT but did not trigger the Fermi/GBM instrument~\citep{2019ApJ...878...52A}. The redshift values are taken from~\url{https://www.mpe.mpg.de/~jcg/grbgen.html}.
\label{tab:Sample1}
\end{table*}

\section{Multi-wavelength Analysis} \label{sect:mwl_analysis}

\subsection{Fermi Gamma Ray Space Telescope}\label{sec:LATanalysis}
The Fermi Gamma-ray Space Telescope consists of two instruments: the Large Area Telescope (LAT) and the Gamma-ray Burst Monitor (GBM). LAT is a pair-conversion telescope comprising a 4$\times$4 array of silicon strip trackers and cesium iodide (CsI) calorimeters, shielded by a segmented anti-coincidence detector to suppress charged-particle background events. It detects gamma-rays from 30 MeV to more than 300 GeV with a 2.4 sr field of view, scanning the entire sky every 3 hours in survey mode \citep{2009ApJ...697.1071A}. 

For time-resolved spectral analysis, we used the \textit{gtburst}\footnote{\url{https://fermi.gsfc.nasa.gov/ssc/data/analysis/scitools/gtburst.html}} tool for extraction and processing of LAT data. We defined a circular region of interest (ROI) of radius $12^\circ$ centered on each burst position and performed standard unbinned likelihood analysis within the first 10\,ks post-BAT trigger. 
We used the \texttt{P8R3\_TRANSIENT020} event class, suitable for transient-source analysis, and the corresponding instrument response functions. As the spectral model, particle background, and Galactic component we assume \texttt{powerlaw2}, \texttt{isotr template}, and \texttt{template (fixed norm.)} respectively. We considered a minimum test statistic (TS\textsubscript{min}) of 20 which denotes the detection significance ($\sqrt{\rm TS}$) of more than 4$\sigma$. For each time bin in which the GRB was within the FoV of LAT, we calculated the 2$\sigma$ (95\% confidence level) upper limits (U.L.) for non-detections (TS$<$20) and the observed flux with 68\% errors (TS$>$20).

We performed a dedicated search for highest energy photons up to energy of 100 GeV with more than 90\% probability of association with the transient in each time interval using the \textit{gtsrcprob} module. 
The flux (or U.L.) and indices ($\Gamma_{\gamma}$)  are reported in 0.1–10 GeV.
An additional unbinned analysis was performed in the energy range between 0.1 GeV and the highest energy photon (E$_{\rm max}$) to estimate the flux (F$_{0.1\,{\rm GeV}-{\rm E}_{\rm max}}$) and flux U.L. (in case of a non-detection; TS$<$20) and the spectral index ($\Gamma'_{\gamma}$). Limiting the analysis to energy bins up to E$_{\rm max}$ reduces systematic uncertainties in the calculation of flux that could arise from extending the energy range beyond E$_{\rm max}$. In addition, we also report flux at 1~GeV for comparison at specific energies. This value is derived from the power-law (PL) fit performed over the 0.1–10 GeV energy range.

\subsection{Neil Gehrels Swift Observatory}
The Neil Gehrels Swift Observatory \citep{SwiftScience:2004ykd} mission was launched in 2004 and has three instruments onboard that allow for multi-wavelength observations across hard and soft X-ray, ultraviolet, and optical wavebands: the Burst Alert Telescope \citep[BAT;][]{Barthelmy:2005hs}, the X-ray Telescope \citep[XRT;][]{SWIFT:2005ngz}, and the UltraViolet Optical Telescope \citep[UVOT;][]{Roming:2005hv}. The analysis techniques for XRT and BAT are described below. 

\subsubsection{Swift/XRT}\label{sec:XRTanalysis}
The XRT is a grazing-incidence-focusing X-ray telescope, covering an energy range of 0.3--10~keV. For the study, data are obtained from the Swift Science Data Center supported by the University of Leicester~\citep{Evans:2008wp}, and include exposures taken in both Window Timing (WT) and Photon Counting (PC) modes.

For the analysis, we use the \texttt{XSPEC} software package (v12.15.0), applying the C-statistic for parameter estimation. For all GRBs, except GRB160325A\footnote{The XRT spectral analysis for GRB160325A was performed in the 1–10~keV energy range due to strong absorption below 1\,keV.}, we fit the time-resolved spectra in the 0.3--10~keV band with a simple PL model, accounting for Galactic absorption using the multiplicative \texttt{tbabs} model. We adopt the Galactic neutral hydrogen column densities from the  GRB spectrum repository of XRT~\citep{Evans:2008wp}. We also apply the \texttt{ztbabs} model to account for intrinsic absorption in the host galaxy, with the host galaxy hydrogen column density ($N_H(z)$) left as a free parameter.
We report the best-fit parameters, unabsorbed flux (F$_{0.3-10~\mathrm{keV}}$), and photon index ($\Gamma_{\rm X}$) in Tab.~\ref{tab:timeresolved}\footnote{Tab.~\ref{tab:timeresolved} presents the time-resolved spectral analysis results for GRBs which includes XRT (0.3–10~keV) and LAT (0.1–10~GeV and 0.1~GeV–E$_{\rm max}$).The final column indicates whether a BAT observation was recorded during the respective time interval.}. In addition, we also report flux at 1~keV for comparison. This value is derived from the PL fit performed over the 0.3–10 keV energy range.

For GRBs with unknown redshift, we adopt $z=0$ when fitting the X-ray spectra~~\citep{Evans:2008wp}. In addition, we present a comparison between the estimated unabsorbed flux, spectral index, $N_H(z)$, and fit statistics for the cases of $z=0$ and $z=2$ (see Tab.~\ref{tab:nh_stats_unknownz}). The fit statistics and $N_H(z)$ for known redshift GRBs are reported in Tab.~\ref{tab:XRT_knownz}.

\subsubsection{Swift/BAT} \label{sec:BATanalysis}
The BAT is a wide-field coded-mask telescope operating in the 15--150~keV energy range with a 1.4 sr field of view. The coded mask technique of BAT provides rapid positional accuracy of 1--4 arcminutes, typically within seconds of a GRB trigger. 
The selected time intervals occur after the BAT T$_{90}$; under this assumption, any post-T$_{90}$ BAT emission is treated as part of the early afterglow.

We obtained hard X-ray data (15–150~keV) from the BAT archive, and processed them with the \texttt{batgrbproduct} pipeline. We generated the spectra using the \texttt{batbinevt} task and included systematic errors with the \texttt{batupdatephakw} and \texttt{batphasyserr} commands. We produced the response matrix with the \texttt{batdrmgen} task. Using \texttt{XSPEC} (v12.15.0), we verified the exposure time of each spectrum to confirm the data availability.

We performed a time-resolved spectral analysis with \texttt{XSPEC} using $\chi^2$ statistics. We fitted the spectra using a simple PL model. 
The resulting best-fit parameters, fluxes, and the photon indices ($\Gamma_{\rm B}$) are reported in Tab.~\ref{tab:bat_results}. For non-detections, we report a flux U.L. with a 90~\% confidence level.

\subsubsection{Joint Swift XRT and BAT}\label{sec:XRTBATjoint}
We performed joint spectral analysis using data from XRT and BAT to constrain the  spectral component at lower energies. For the joint fitting, we used $\chi^2$ statistics for the BAT data and C-statistics for the XRT data. 
We modelled the spectra in the 0.3--150~keV energy band using two functions: PL and smoothly broken power law (SBPL). 
The SBPL is defined as:
\begin{equation}
N_E = A 
\left[
\left( \frac{E}{E_j} \right)^{-\alpha n}
+
\left( \frac{E}{E_j} \right)^{-\beta n}
\right]^{-\frac{1}{n}},
\end{equation}
where
\begin{equation}
E_j = E_p 
\left(-
\frac{\alpha + 2}{\beta + 2}
\right)^{\frac{1}{(\beta - \alpha)n}}.
\end{equation}

Here, $N_E$ represents the photon spectrum, $A$ is the normalization, $E_p$ is the peak energy, $\alpha$ is the spectral slope below the peak,
$\beta$ is the spectral slope above the peak, and $n$ is the smoothness
parameter, which is fixed to $n = 2$. While the PL serves as a baseline model, the SBPL captures a potential spectral peak. 
 
For the joint spectra, we accounted for Galactic absorption using the \texttt{tbabs} model, based on the estimated neutral hydrogen column density along the line of sight. Intrinsic host-galaxy absorption is modeled with \texttt{ztbabs}, with the redshift-dependent host-galaxy column density $N_{\rm H}(z)$ as a free parameter.

\begin{figure}[h]
    \centering
    \includegraphics[width=0.95\columnwidth]{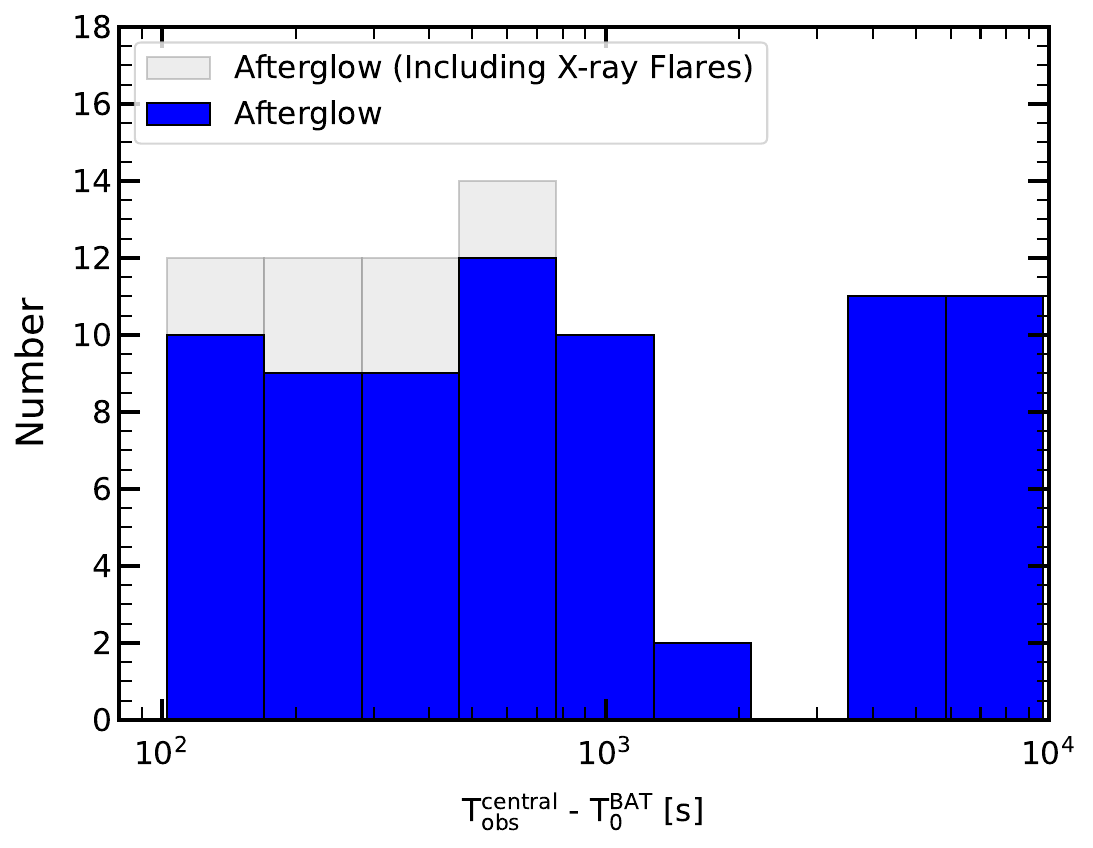}
    \caption{Distribution of time-bin centers relative to BAT trigger time. The histograms show the distribution of time bins centers with respect to the BAT trigger time. Time bins exhibiting consistent temporal decay behavior in afterglow are shown in blue, while those containing X-ray flares are shown in gray.}
    \label{fig:Time-bin}
\end{figure}

\section{Time-bin Selection}\label{sec:Time-bin selection}
In our study, the time windows that have simultaneous observation with XRT and LAT are selected based on the following criteria:
\begin{itemize}
    \item Observation modes of XRT (WT or PC, see Sect.~\ref{sec:XRTanalysis})
    \item The time window of source inside the FoV of LAT satisfying the observational constraints: zenith angle $<$ 100$^{\circ}$ and source off-axis angle $<$ 65$^{\circ}$.
\end{itemize}

In addition, a time-window is further subdivided on the basis of the temporal break(s) identified by the XRT light curve modeling pipeline \citep{2009MNRAS.397.1177E}. This results in 85 time bins among 31 GRBs. Among them, 11 time bins show evidence of X-ray flares\footnote{A complete list of these time bins with X-ray flares (identified using the Swift automated pipeline~\citep{Willingale:2006zh}) are reported in Tab.~\ref{tab:Flare_GRBs}. All flares occur within 1000 seconds after the burst.}. The study related to the emission mechanism of X-ray flares is beyond the scope of this work and will be discussed in future. Therefore, we compile the list presented in Tab.~\ref{tab:timeresolved} that reports 74 time bins excluding X-ray flares. The last column of Tab.~\ref{tab:timeresolved} reports the availability of the BAT data. Furthermore, we identified 34 bins having simultaneous XRT, BAT, and LAT observations.

The distribution of the central time bin (T$^{\rm central}_{\rm obs}$) relative to the BAT trigger time is reported in Fig.~\ref{fig:Time-bin}. The gray and blue histograms represent the afterglow emission, including and excluding the flares, respectively. The distribution exhibits a gap between approximately 1.5\,ks and 5\,ks. This gap arises because the analyzed time intervals require the GRB to be within the LAT's field of view, and LAT observes any given point in the sky with a cadence of about 90\, mins ($\sim$5.5\,ks).

\begin{figure*}[ht!]
    \centering
     \includegraphics[width=2.0\columnwidth]
     {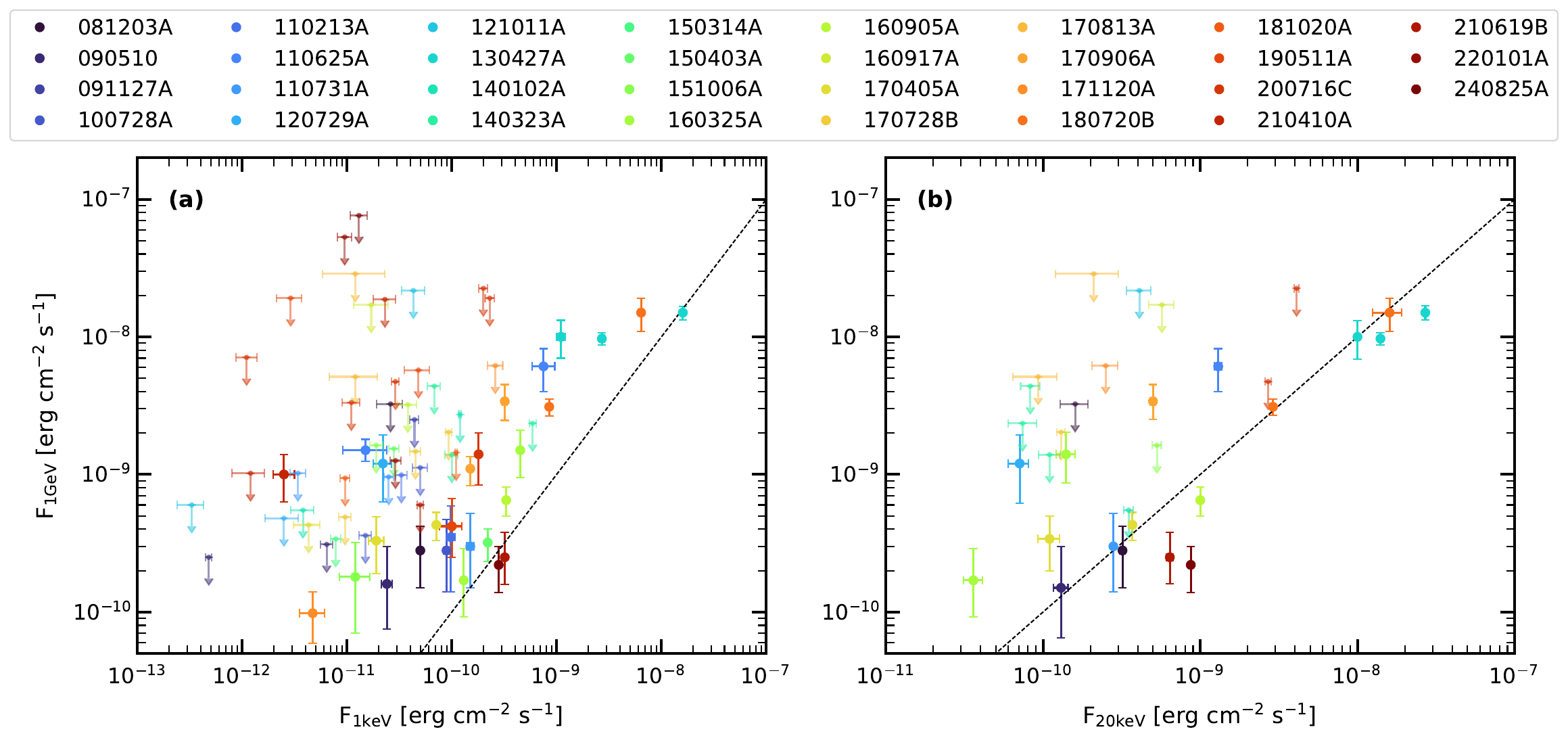}
     \caption{Comparison between X-ray and HE gamma-ray fluxes. Plot (a) and (b) presents comparison of X-ray and HE gamma-ray fluxes. \textbf{(a)}: Flux in 1~keV vs. 1~GeV.  \textbf{(b)}: Flux in 20~keV vs. 1~GeV. Each GRB has assigned unique color consistent throughout the paper. The dashed line indicates equality line. Both panels show LAT detections (data points with flux uncertainity, \(\text{TS} > 20\)) and upper limits (downward-pointing with reduced opacity). The right panel has fewer bins than left panel due to limited sensitivity of BAT (see Tab.~\ref{tab:timeresolved} and Sect.~\ref{sec:flux_flux} for details).}
    \label{fig:flux_XRT_LAT}
\end{figure*}

\section{Results} \label{sec:results}
In order to characterize the broad band afterglow emission and extract relevant information on the afterglow microphysical parameters, we used the observable properties, such as fluxes and photon indices in X-ray and HE gamma-rays. In particular, we considered two representative energies for calculating flux in the X-ray band (1 and 20~keV) and one representative energy in the HE gamma-ray band (1~GeV).

\subsection{Comparison between X-ray and HE gamma-ray flux} \label{sec:flux_flux}
In this section, we compare the HE gamma-ray flux at 1~GeV (F$_{1~\mathrm{GeV}}$) with the X-ray fluxes at two distinct energies, 1~keV (F$_{1~\mathrm{keV}}$; representing XRT energy band) and 20~keV (F$_{20~\mathrm{keV}}$; representing joint XRT and BAT energy band). We note that data points in XRT are more numerous than joint XRT and BAT due to the limited detections in BAT.

\subsubsection{Soft X-ray (F$_{1~\mathrm{keV}}$) vs. HE gamma-ray (F$_{1~\mathrm{GeV}}$)}\label{sec:XRTLAT_flux}
We investigated the flux–flux correlation between the X-ray band at 1 keV and the HE gamma-ray band at 1 GeV. Figure~\ref{fig:flux_XRT_LAT} (a) shows the correlation between the XRT flux at 1 keV (F$_{1~\mathrm{keV}}$) and the LAT flux at 1 GeV (F$_{1~\mathrm{GeV}}$), as listed in Table~\ref{tab:timeresolved}. We note that flux at 1 GeV is systematically higher than 1 keV.

In addition, we further compared fluxes normalized by the prompt emission fluence ($S_{\gamma}$; 1-1000\,keV) of each burst observed with Fermi/GBM as reported in Tab.~\ref{tab:Sample1}. We note that these normalized fluxes (see Fig.~\ref{fig:flux_normalized_XRT_LAT}(a)) in the X-ray and GeV energy bands clearly shows a systematic departure from unity. 

\subsubsection{X-ray (F$_{20~\mathrm{keV}}$) vs. HE gamma-ray (F$_{1~\mathrm{GeV}}$)}\label{sec:XBLAT_flux}
We performed a joint spectral analysis with XRT and BAT for the time bins depending on the availability of the hard X-ray data from BAT. The majority of the spectra (28 out of 34; see Sect.~\ref{sec:Time-bin selection}) show preference for a PL spectrum. Among these 28 spectra, in 23 cases, we are limited to XRT observations only, with BAT providing  upper-limits (see Tab.~\ref{tab:bat_results}). Hence, a curvature in the spectra could not be estimated, resulting in a preference for a PL model. For BAT upper limit cases, the joint XRT and BAT fit is effectively a straightforward extrapolation of the XRT spectrum into the BAT energy range, with the BAT upper limit imposing a ceiling on the spectral shape. 
The best-fit parameters, including the unabsorbed fluxes (F$_{0.3-150~\mathrm{keV}}$), photon indices ($\alpha$), and flux at 20~keV (F$_{20~\mathrm{keV}}$) are reported in Tab.~\ref{tab:xrt_bat_PL}.
 
In a total of 6 cases (out of 34), the SBPL provides significantly better fit than PL model. To obtain the best model, we compare the total statistics (sum of $\chi^2$ and C-statistic) per degree of freedom (see Tab.~\ref{tab:Pl_sbpl} for details). The SBPL model provides an estimate of the peaks between 0.3-150~keV. The best-fit SBPL parameters including the unabsorbed fluxes (F$_{0.3-150~\mathrm{keV}}$), peak energy (E$_{\mathrm{p}}$), low-energy photon indices ($\alpha$), high-energy photon indices ($\beta$), and flux at 20~keV (F$_{20~\mathrm{keV}}$)  are reported in Tab.~\ref{tab:xrt_bat_SBPL}. 
Figure~\ref{fig:flux_XRT_LAT} (b) shows a comparison between the flux measured at 20~keV and at 1~GeV. Although the total number of observations is smaller than that the reported in Sect.~\ref{sec:XRTLAT_flux}, the flux values are symetrically distributed around the equality line. In addition, a correlation emerges when the fluxes in both X-rays (20~keV) and HE gamma-rays (1~GeV) are normalized by the $S_{\gamma}$ (1–1000~keV; see Tab.~\ref{tab:Sample1} and Fig.~\ref{fig:flux_normalized_XRT_LAT}(d)).

\begin{figure}[h]
    \centering
    \includegraphics[width=\linewidth]{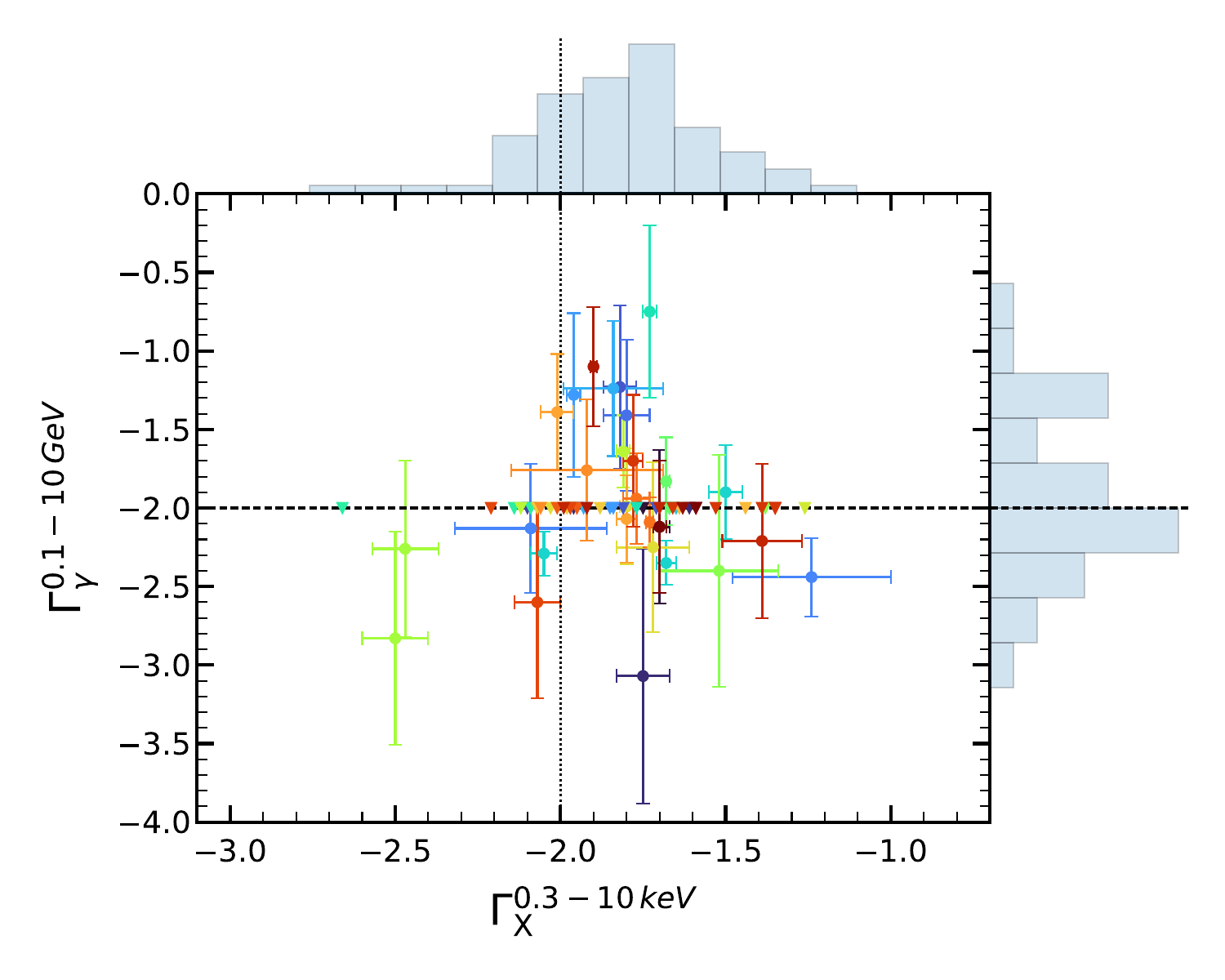}
    \caption{X-ray vs HE gamma-rays photon indices. Comparison of photon indices between X-rays (0.3--10 keV; x-axis) and HE gamma rays (0.1--10 GeV; y-axis). The histograms shown on the top and right panels display the distributions of photon indices in the XRT and LAT bands, respectively. Two reference lines are plotted: one at a LAT photon index of $-2$ and one at an XRT photon index of $-2$. Data points with error bars correspond to significant LAT detections ($\mathrm{TS} > 20$), while upper limits (calculated in the 0.1--10 GeV band) are indicated by downward arrows ($\mathrm{TS} < 20$). For further details, see Table~\ref{tab:timeresolved} and Section~\ref{section:Index comparison}.}
\label{fig:Index_LAT_XRT}
\end{figure}

\subsection{Comparison between spectral indices in X-rays and GeV bands} \label{section:Index comparison}
Comparing the photon indices across X-ray and GeV energies provides a powerful diagnostic of the dominant radiation mechanisms governing the GRB afterglow. By examining the correlation between $\Gamma_{\rm X}$ and $\Gamma_\gamma$, we can test whether the broadband emission arises from a single synchrotron component or whether an additional spectral component (such as SSC) is required.

To evaluate the relative spectral behavior, we compared the photon indices in the X-ray band ($\Gamma_{\rm X}$; 0.3-10~keV) and HE gamma-rays ($\Gamma_\gamma$; 0.1-10~GeV) reported in Tab.~\ref{tab:timeresolved}. Only one time bin of GRB130427A is identified with the highest energy photon above 10 GeV (57~GeV, see Tab.\ref{tab:Sample1}). The choice of a unique energy bin of 0.1-10\,GeV is thus justified to further compare the results with the theoretical prediction. Figure~\ref{fig:Index_LAT_XRT} shows the index comparison between X-rays and HE gamma-rays. 
In addition, we investigated the index comparison between the X-rays and the dynamical energy range in GeV, namely $\Gamma'_\gamma$; 0.1\,GeV-E$_{\rm max}$. We note that the spectral index estimated up to E$_{\rm max}$ is systematically harder than the index estimated in the energy range of 0.1-10\,GeV (see Fig.~\ref{fig:Index_LAT_comparison} and Fig.~\ref{fig:Index_EmaxLAT_XRTcomparison} for details).  

In Fig.~\ref{fig:Index_LAT_XRT}, we observe that the data points populate three broad regions of the XRT–LAT photon index plane. The first region corresponds to harder X-ray emission in 0.3-10~keV and is associated with softer GeV emission in the 0.1--10~GeV band ($\Gamma^{\rm 0.1-10GeV}_{\gamma}$<-2.0). The typical spectra (39 cases out of 74) are possibly represented by a single broadband (0.3\,keV-10\,GeV) spectral component. GRB 130427A belongs to this spectral class in the time-bin 250-463\,s after the BAT trigger (see Fig.~\ref{fig:130427A_SED} for details). The second region corresponds to moderately harder X-ray emission and is linked with harder GeV emission ($\Gamma^{\rm 0.1-10GeV}_{\gamma}>$-2.0). The spectra (19 cases out of 74) the spectra potentially indicate a double spectral component, requiring an additional spectral component in addition to the synchrotron emission. GRB 180720B, in the time window 147-625s after the BAT trigger, belongs to this case.  Fig.~\ref{fig:180720B_SED} 
demonstrates the potential presence of two distinct spectral components in GRB 180720B, as indicated by harder X-ray and GeV spectra. The third region corresponds to X-ray photon indices less than $-2.0$, representing the softer emission in 0.3-10~keV (16 cases out of 74). The detection in GeV energies with softer X-ray spectra indicates the presence of an additional spectral component. GRB 160325A is one of the examples for this category (see Fig~\ref{fig:160325A_SED} for details) with softer spectral index in X-rays during 230-504\,s after the BAT trigger.
\begin{figure*}[h]
    \centering
    
    \includegraphics[width=2\columnwidth]{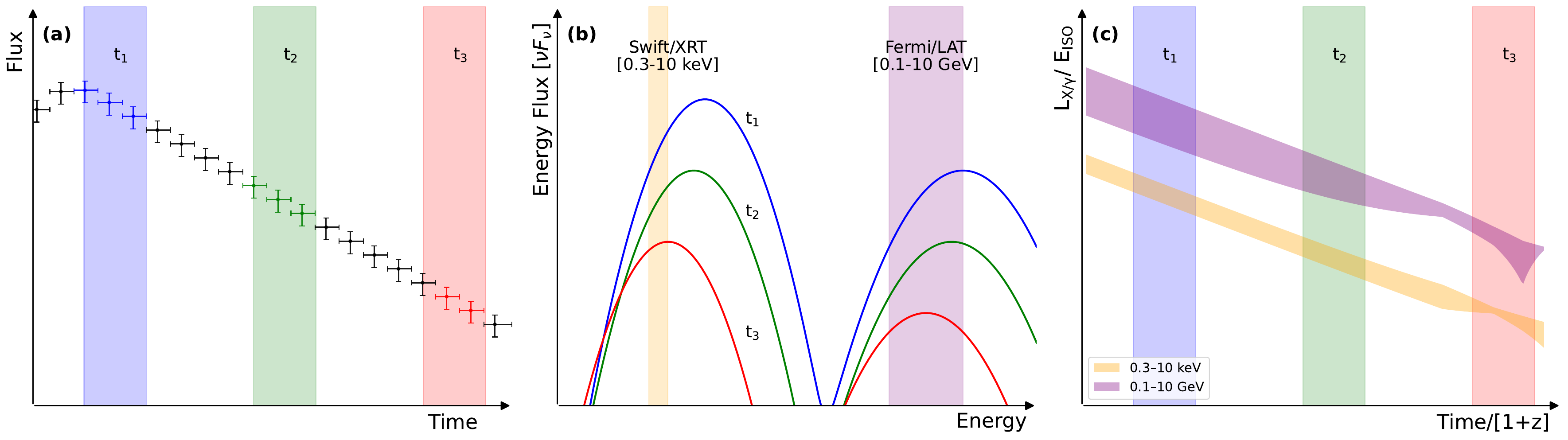}
    \caption{Illustration of the spectral and time evolution of the afterglow emission. \textbf{(a)} typical light curve in arbitrary energy band of a GRB afterglow with three temporal bins highlighted in blue, green, and red. Plot \textbf{(b)} Spectra corresponding to the three temporal bins from panel (a). The orange and purple shaded regions indicate the energy ranges covered by Swift/XRT (0.3–10 keV) and Fermi/LAT (0.1–10 GeV), respectively. \textbf{(c)} Relation between the luminosity in X-ray ($L_{\rm X}$) and HE gamma-rays ($L_{\rm \gamma}$) and the prompt emission isotropic energy ($E_{\rm iso}$) in the rest frame of GRBs, shown for the energy ranges 0.3–10 keV (orange) and 0.1–10 GeV (purple). The x- and y-axes in all panels are in arbitrary units.}
    \label{fig:illustration}
\end{figure*}

\section{Inference of microphysical parameters and application in present work} \label{sec:Intrepretation}
\subsection{Estimation of microphysical parameters from afterglow correlations}
Previous independent data-driven studies have aimed to establish a unique temporal (in the rest frame) decline of the ratios $L_{\rm X}/E_{\rm iso}$ and $L_{\gamma}/E_{\rm iso}$ during the GRB afterglow phase \citep{2012MNRAS.425..506D, 2014MNRAS.443.3578N, 2019ApJ...878...52A}. 
In this context, $L_{\rm X}$ and $L_{\gamma}$ denote the luminosities measured in the X-ray and gamma-ray energy bands, respectively, and $E_{\rm iso}$ is the isotropic-equivalent total radiated energy.

This section presents a methodology for identifying a unique parameter set describing afterglow emission capable of reproducing these observed temporal behaviors through physical modeling. We subsequently derive the allowed flux (in X-ray and GeV) and spectral index contours from these preferred parameters. We compare these predictions with our multi-wavelength dataset described in Sect.\ref{sec:results}.

Figure~\ref{fig:illustration} shows a representation of the spectral evolution of a standard forward shock GRB afterglow. Panel (a) shows the light curve of the afterglow with a constant temporal decay and three highlighted time bins. Panel (b) displays the spectra corresponding to these time bins and energy ranges of Swift/XRT (0.3–10 keV) and Fermi/LAT (0.1–10 GeV). Panel (c) presents the relation between isotropic luminosity and prompt emission isotropic energy in the rest frame of GRBs for the energy ranges detectable by 0.3–10 keV and 0.1–10 GeV.

\subsubsection{Model description}
We used the leptonic module of the Lepto-Hadronic Modeling Code~\citep[LeHaMoC;][]{Stathopoulos:2023qoy}, called LeMoC, to model our multi-wavelength dataset. LeMoC follows the evolution of relativistic electron populations interacting with magnetic and photon fields within a spherical region, while accounting for synchrotron radiation, synchrotron self-absorption, inverse Compton (IC) scattering, photon-photon absorption, and adiabatic losses. In this work, we adapt the code to model GRB afterglows under simplified assumptions. We consider the afterglow emission to be dominated by synchrotron and IC radiation produced by shock-accelerated electrons following a PL energy distribution, expressed as $dN/d\gamma \propto \gamma_e^{-p}$, where $p$ denotes the electron spectral index and $\gamma$ represents the electron Lorentz factor. At the start of the simulation, electrons are injected once with the electron spectral index $p$ into an expanding spherical region of initial radius $R_{e}$. The comoving magnetic field $B$ is assumed to remain constant throughout the evolution. We further assume that the fraction of the total energy in the shocked region that accelerates electrons into a PL distribution, $\epsilon_{\rm e}$, does not vary with time. The minimum and maximum Lorentz factors of the injected electrons are given by $\gamma_m$ and
$\gamma_{\text{max}} \approx \left( \frac{6 \pi q_e}{\sigma_T B} \right)^{1/2}$, where $q_e$ is the electron charge and $\sigma_T$ is the Thomson cross section, respectively. This setup enables the computation of broadband spectra at different epochs by exploiting the self-similar dynamics of relativistic blast waves in a cold medium~\citep{1976PhFl...19.1130B}. 

In order to produce a spectrum, the following five parameters are varied: the index of the electron energy distribution $p$, the equipartition parameters such as fraction of shocked energy goes to electrons $\epsilon_{\rm e}$, the fraction of energy to the magnetic field $\epsilon_{\rm B}$, prompt emission efficiency $\eta$ and density of medium. We considered both homogeneous and wind medium scenarios. In case of the wind-medium, we use general form of the density $n(R)$ that can be expressed as A$\times\rm R^{-2}$ where A is given by A$_{*}\times10^{35}$ cm$^{-1}$ (A$_{*}$ is dimensionless normalization factor). For the homogeneous medium we use a constant density n. 

\begin{figure}[t]
    \centering
    \includegraphics[width=\columnwidth, height=6cm]{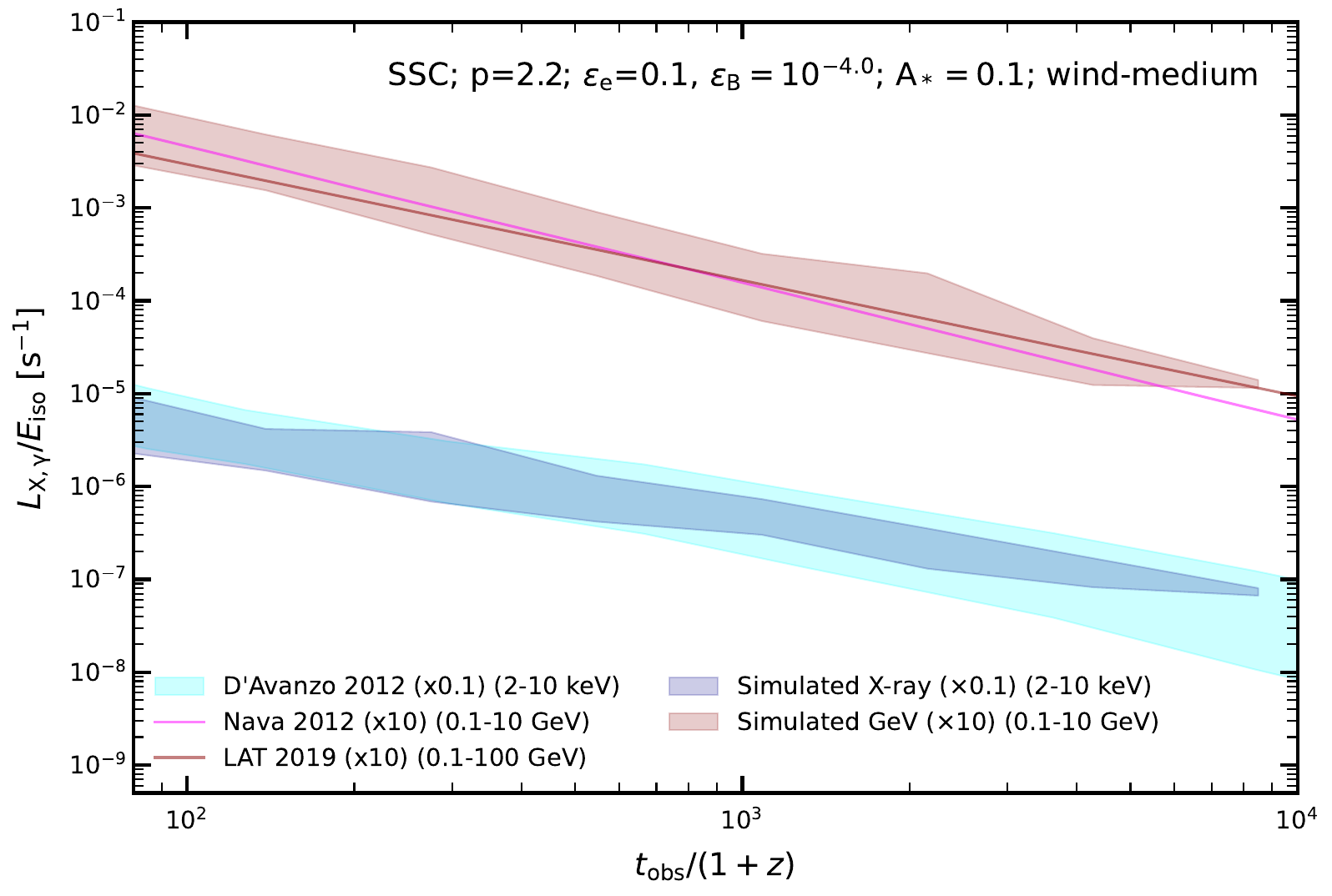}
    \caption{The relation between the luminosity in X-ray ($L_{\rm X}$)  and HE gamma-rays ($L_{\gamma}$) with the prompt emission isotropic energy (E$_{\rm iso}$) as a function of the rest frame time of GRBs. The solid pink and magenta lines represent the trends reported by~\citealt{2014MNRAS.443.3578N} and \citealt{2019ApJ...878...52A}, respectively, for the ratio L$_\gamma/E_{\rm iso}$. The shaded cyan band indicates the $1\sigma$ confidence region for the corresponding trend in X-rays (L$_{\rm X}/E_{\rm iso}$) reported in \citealt{2012MNRAS.425..506D}, where $L_{\rm X}$ is calculated in the 2--10\,keV band. The magenta and blue band represents the simulated GeV and X-ray emission for the SSC parameters \(p=2.2\), \(\epsilon_{\rm e} = 0.1\), \(\epsilon_{\rm B} = 10^{-4}\) in wind medium with A$_{*}$ = 0.1. }

    \label{fig:LeMoC_1}
\end{figure}

\subsubsection{Simulation setup} \label{sec:simulation_setup}
We test the hypothesis that afterglow emission can be described by a unique set of parameters: $p$, $\epsilon_{\rm B}$, medium: homogeneous or wind and density of medium. We simulate a number of realizations, where for each, we chose one value of $p$ from \{2.2, 2.3, and 2.4\} (3 combinations), one value of $\epsilon_{\rm B}$ from \{10$^{-2}$, 10$^{-3}$, and 10$^{-4}$\} (3 combinations). Each unique combination of these parameters $(p,\epsilon_{\rm B},\text{medium}, \text{density})$ defines what we refer to as a "benchmark". 
In addition, for two different density profiles, two density values are used: for the homogeneous medium, we use [1 cm$^{-3}$, 0.1 cm$^{-3}$], and [A$_{*}$=0.1, 1] for wind-medium. 
In total, we construct 36 distinct benchmarks (three values of $p \, \times$ three values of $\epsilon_{\rm B} \, \times$ two types of medium with two values of densities), each representing a different scenario for afterglow emission.
 
For each benchmark, we fix the prompt emission efficiency $\eta=0.1$ (\citealt{2002A&A...391.1141D}, \citealt{2003astro.ph.12347L}, \citealt{2005A&A...430....1G}, and \citealt{2008A&A...480..305G})  and $\epsilon_e=0.1$\footnote{We also tested $\epsilon_e = 0.02$ and $\epsilon_e = 0.5$, but neither value is preferred.}
 (\citealt{2012MNRAS.425..506D, 2015MNRAS.454.1073B, 2022MNRAS.511.2848A}). These parameter values serve as practical working assumptions, since these microphysical and dynamical quantities are degenerate. We sample 220 random values of E$_{\rm ISO}$, z and observation time up to 10$^{4}$\,s that mimics the observation of one year of long GRBs\footnote{There is only one short GRB in our sample: GRB090510. Hence, we only consider long GRBs for the simulations.}. The values of redshifts (z) and E$_{\rm ISO}$ are selected in such a way that they follow the distributions reported in \citealt{2022arXiv220606390G}. Moreover, the selected GRBs are above the detectability limit of GBM in z vs. E$_{\rm ISO}$ plane (see Fig.~\ref{fig:Eiso_z} in the Appendix). The detectability limit has been adapted from \citealt{2023ApJ...952L..42L}.
Thus, for each benchmark, we construct a total of 220 simulated random afterglow spectra (representing individual GRBs; see Sect. \ref{sec:sims} for details) in X-ray (0.3-10~keV) and HE gamma-rays (0.1-10 GeV) at any random time within 10\,ks from the trigger. For the simulated afterglow at different observed times (t$_{\rm obs}$), the bulk-Lorentz factor ($\Gamma$) is always below 300 (which may be considered as a typical initial bulk-Lorentz factor, $\Gamma_0$), confirming that the simulated afterglow is always in the deceleration phase.

\subsubsection{Preferred model parameters}\label{sec:preferred_model}
The 220 realizations provide the integrated fluxes and the photon indices in X-rays (0.3-10 keV), HE gamma-rays (0.1-10 GeV), and VHE gamma-rays (0.3-1 TeV)\footnote{The flux in VHE gamma-rays is an intrinsic flux which is defined as the flux emitted at the source, without being attenuated due to EBL.}. 
Although the flux and the spectral index derived for the X-ray are in 0.3-10\,keV, the trend mentioned in~\citealt{2012MNRAS.425..506D} is between 2-10 keV. Following the same approach as~\citealt{2012MNRAS.425..506D}, we compute the flux in 2-10 keV (F$_{\rm 2-10\,keV}$) from the integral flux in the energy range of 0.3-10 keV (F$_{\rm 0.3-10\,keV}$) and the photon index ($\Gamma_{\rm X}$) as follows:

\begin{equation}
    {\rm F}_{\rm 2-10\,keV} = {\rm F}_{\rm 0.3-10\,keV} \times \left[ \frac{1-\left(2/10\right)^{2+\Gamma_{\rm X}}}{1-\left(0.3/10\right)^{2+\Gamma_{\rm X}}}
    \right]
\end{equation}

Motivated by previous studies~\citep{2012MNRAS.425..506D,2014MNRAS.443.3578N}, we initially explored a set of standard parameters $\{\eta = 0.1,\, \epsilon_{\rm B} = 10^{-2},\, \epsilon_{\rm e} = 0.1\}$ with electron indices $p = 2.2$ and $p = 2.3$ in both wind-like and homogeneous media (see Fig.~\ref{fig:LaraPaolo} in the Appendix). Our analysis reveals that in homogeneous environments, the emission in keV and GeV energies are overproduced for both the cases of $p = 2.2$ and $p = 2.3$.

We additionally calculated the p-value for each benchmark using the KS-test to estimate the similarity between the model and the data driven correlations in X-ray and GeV. Our simulation assesses the preference of reproducing the clustering by comparing the simulated data in GeV and X-rays. This comparison is carried out based on reproducing the GeV clustering with a dispersion of 0.23\footnote{According to \citealt{2014MNRAS.443.3578N}, the dispersion in clustering is represented by $\sigma_{log (L/E)} = 0.23$, indicating the uncertainty in the ratio of luminosity to isotropic equivalent energy during the prompt emission phase for a particular rest frame time.} and within the 1$\sigma$ confidence interval for X-rays from \citealt{2012MNRAS.425..506D}.

Table~\ref{tab:KStest} presents the model parameters along with the p-values for X-ray and GeV, both individually and combined. From the initial choices mentioned in Sect.~\ref{sec:simulation_setup}, depending on KS-test, the most favorable benchmark is the following: $p = 2.2$, $\epsilon_{\rm e} = 0.1$, $\epsilon_{\rm B} = 10^{-4}$ in a wind-like medium with A$_{*}$ = 0.1. 
However, for the case with $p = 2.3$ combined with the same set of parameters $\epsilon_{\rm e}$, $\epsilon_{\rm B}$ and A$_{*}$ mentioned above, there is a similar preference (combined p-value of $>$0.9) for X-ray and GeV clustering (see Tab. ~\ref{tab:KStest}). 
Figure~\ref{fig:LeMoC_1} shows the result of the simulation and the comparison between the trends. The magenta and pink solid lines are representative of the trend of the ratio between the luminosity and the isotropic energy of the prompt emission in the GeV energies for 0.1-10 GeV \citep{2014MNRAS.443.3578N} and 0.1-100 GeV~\citep{2019ApJ...878...52A}, respectively. The cyan band represents the 1$\sigma$ region (L$_{\rm X}$/E$_{\rm ISO}$) in X-rays in the energy band (2-10\,keV)~\citep{2012MNRAS.425..506D}. The simulated quantities are represented by blue and red bands for keV and GeV predictions, respectively.

\begin{figure*}[h!]
    \centering
    \includegraphics[width=2\columnwidth]{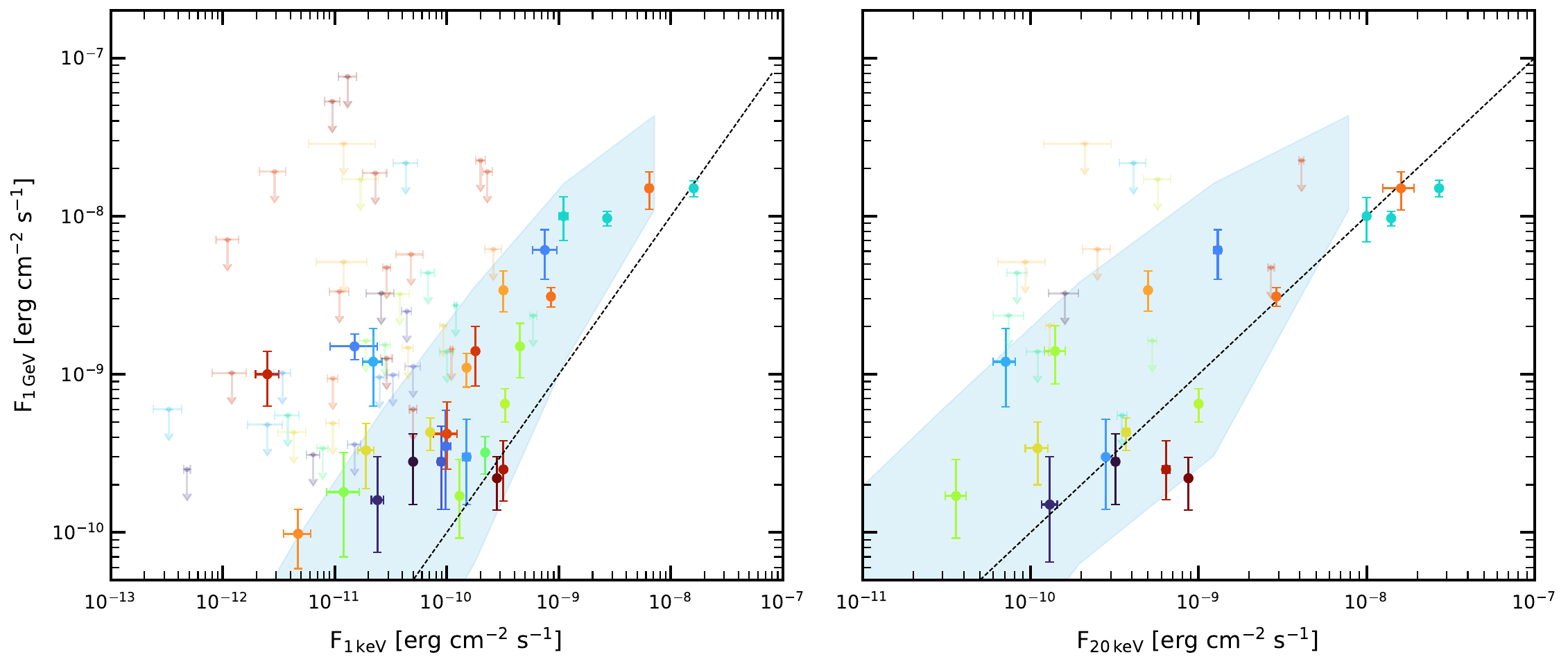}
    \caption{\textbf{Left:} The blue region in  both the \textbf{left} figure and \textbf{right}  figure represents the 1$\sigma$ predicted range of soft X-rays (1 keV), soft and hard X-ray (20 keV) and HE gamma-rays (1GeV). These prediction accounted for synchrotron self-Compton emission in a wind-like medium, modeled with parameters $p = 2.2$, $\epsilon_{\rm e} = 0.1$,  $\epsilon_{\rm B} = 10^{-4}$ and A$_{*}$=0.1. Data points are observed data points present in Tab.~\ref{tab:timeresolved}. 
    }
    \label{fig:LeMoC_2}
\end{figure*}

\subsection{Application to the X-ray and GeV data in this work: prediction of TeV emission}\label{sec:simAfterglow}

The preferred benchmark from the previous section ($p=2.2$, $\epsilon_{\rm e}=0.1$, $\epsilon_{\rm B}=10^{-4}$, wind medium with A$_{*} = 0.1$) is used to explain the observables such as flux and photon indices in X-ray and GeV energy range described in Sect.~\ref{sec:results}. These parameters are determined solely based on previously established empirical correlations in X-rays and GeV energies, as reported in \citep{2012MNRAS.425..506D,2014MNRAS.443.3578N}. As shown in left panel of Fig.~\ref{fig:LeMoC_2}, the predicted region (blue region) aligns well with the observed fluxes at the X-ray band (1\,keV) and the HE gamma-ray band (1\,GeV). Furthermore, this set of microphysical parameters also captures the observed flux trends at higher X-ray energy (20\,keV; see right panel of Fig.~\ref{fig:LeMoC_2}). This agreement across widely separated energies supports a common physical origin for the X-ray and high-energy gamma-ray emission. The fact that most data points lie within the predicted region indicates that the observed GeV (at 1~GeV) and X-ray flux (at 1 keV and 20 KeV) are broadly consistent with the predicted SSC in wind medium.

\begin{figure}[h]
    \centering
    \includegraphics[width=\columnwidth]{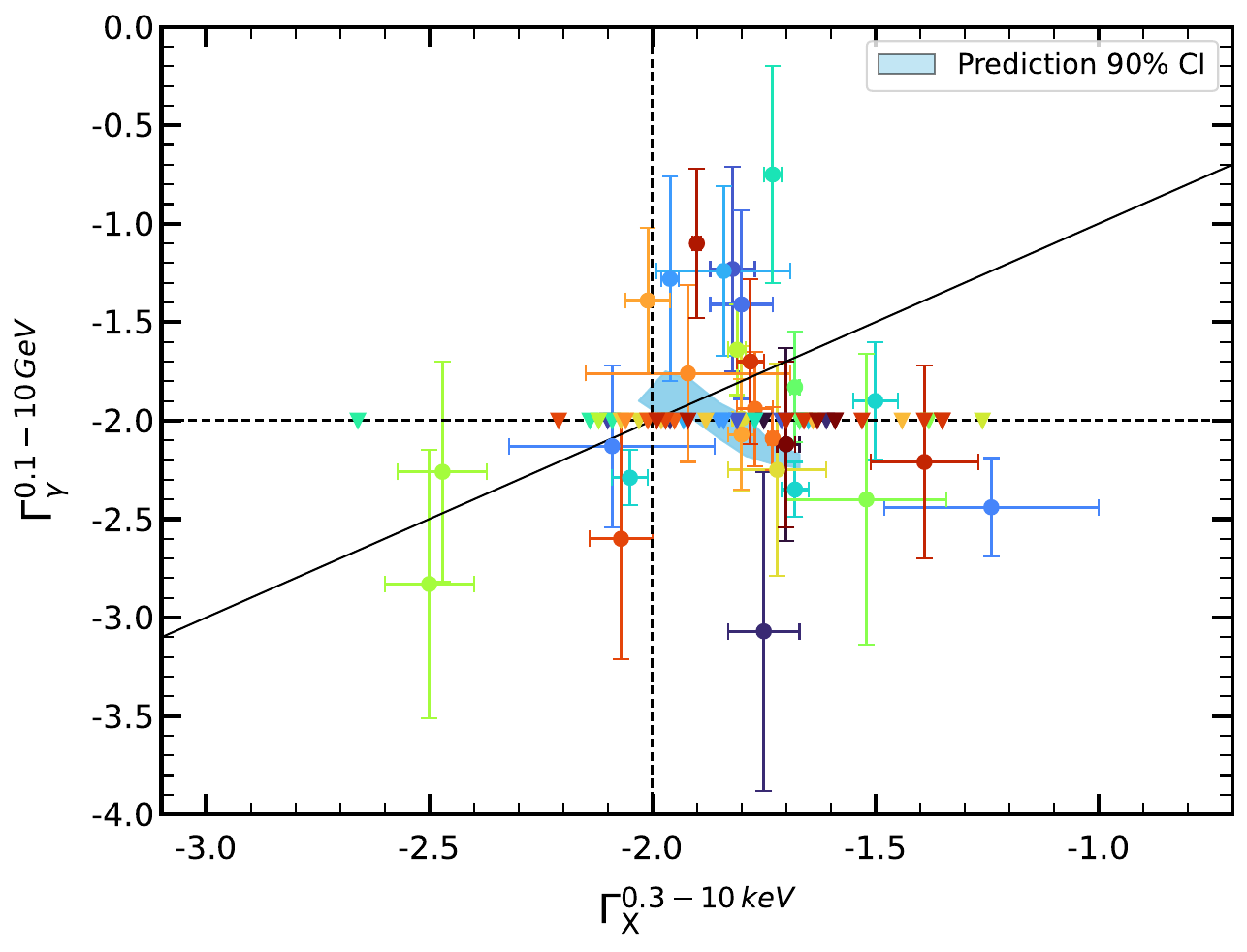}
   \caption{Figure shows the simulated photon indices between X-ray (0.3-10~keV) and HE gamma-rays (0.1-10~GeV) represented in sky-blue region for SSC in wind medium for parameters $p = 2.2$, $\epsilon_{\rm e} = 0.1$, $\epsilon_{\rm B} = 10^{-4}$ and A$_{*}$ = 0.1. The data points are observed photon indices reported in Table~\ref{tab:timeresolved}.}
    \label{fig:LeMoC_3}
\end{figure}

In addition to flux predictions, the parameters reproduce the observed photon indices in the X-ray (0.3–10\,keV) and HE gamma-ray (0.1–10\,GeV) bands. As illustrated in Fig.~\ref{fig:LeMoC_3}, the predicted region shows a clustering of photon indices into two segments, consistent with the observed distributions discussed in Sect.~\ref{section:Index comparison}. The 90\% credibility region naturally incorporates the uncertainties associated with the determination of the GeV photon indices. The observed softer X-ray indices for GRB 160325A can be considered as an outlier.

\begin{figure}[h]
    \centering
    
    \includegraphics[width=\columnwidth]{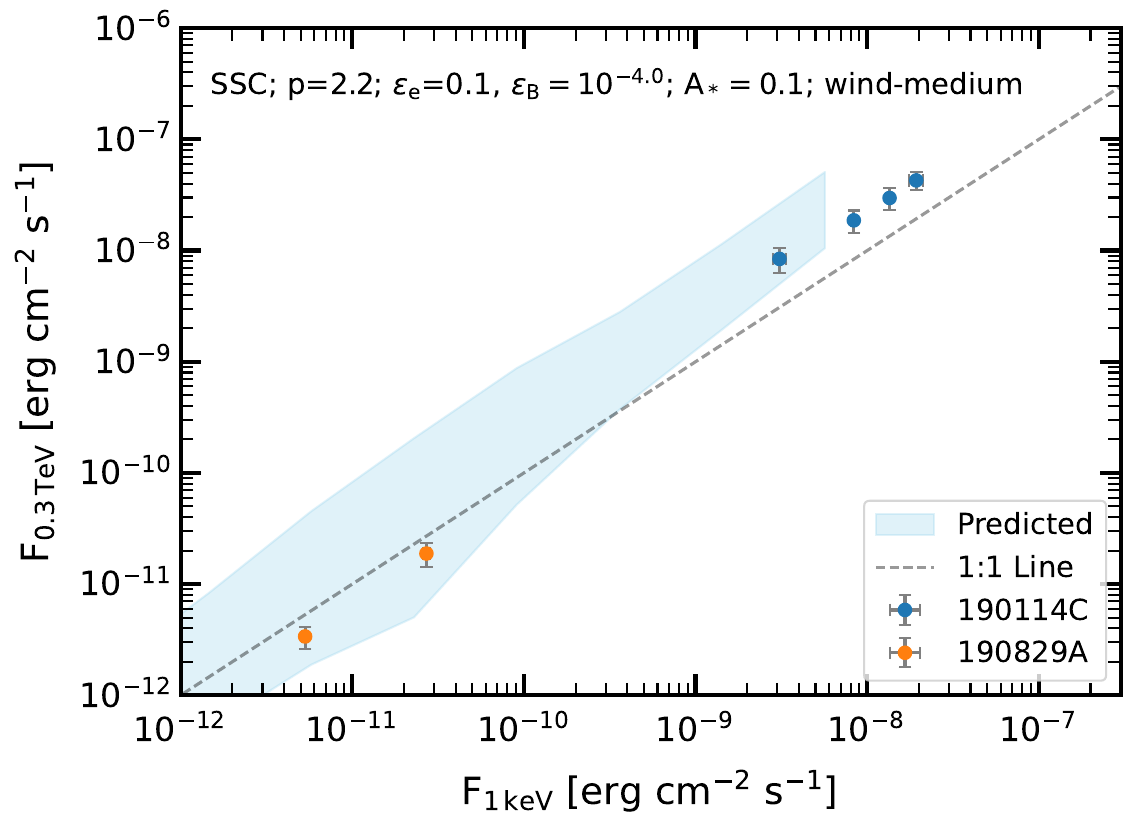}
    \caption{Comparison between predicted intrinsic VHE and X-ray fluxes. The figure presents correlation between the intrinsic VHE flux (0.3~TeV) and the X-ray flux (1~keV), both predicted using SSC emission in a wind-like circumburst medium. The model assumes the parameters $p=2.2$, $\epsilon_{\rm e} = 0.1$,  $\epsilon_{\rm B} = 10^{-4}$ and A$_{*}$ = 0.1. The data points shown by blue and orange color are VHE GRB190114C and GRB190829A respectively (see Appendix~\ref{app:VHE}).
}
    \label{fig:LeMoC_4}
\end{figure}

Finally, we simulate the intrinsic VHE gamma-ray emission at the 0.3 TeV energy range using the preferred benchmark parameters. The resulting intrinsic VHE fluxes exhibit a clear positive correlation with the X-ray flux at 1 keV (see Fig.~\ref{fig:LeMoC_4}). Such a trend is naturally expected if both emission components originate from the same population of relativistic electrons, producing synchrotron radiation in the X-ray band and inverse Compton emission in the VHE regime. This predicted correlation therefore establishes a direct and testable connection between X-ray afterglow brightness and TeV detectability. For test, we also include the intrinsic VHE flux for GRB~190114C and GRB~190829A, shown in Fig.~\ref{fig:LeMoC_4} as blue and orange data points, respectively (see Appendix~\ref{app:VHE}). Since the predicted VHE flux was computed assuming a $\Gamma$ < 300 and avoiding prompt contamination, it does not account for GRB~190114C ($\Gamma$ > 500; \citealt{MAGIC:2019lau}). The rest of the events are broadly consistent with the model-predicted correlation, lending additional observational support to the SSC interpretation in a wind-like circumburst environment.

\section{Discussion and summary} \label{Discussion}
In this paper, we use multiwavelength observations from X-ray and GeV detectors to investigate the emission processes during the early afterglow phase of GRBs. These instruments collectively enable us to probe the emission over a broad energy range, spanning more than seven orders of magnitude from soft X-rays at 0.3~keV up to HE gamma rays above 100 GeV. Our analysis aims to refine estimates of key microphysical afterglow parameters, including the magnetic field strength, the fraction of energy imparted to electrons, and the electron energy distribution, in order to simultaneously explain both the X-ray and GeV emissions. Additionally, we seek to characterize the circumburst environment surrounding the progenitor, which plays an important role in shaping the observed HE and VHE emission.

We selected a sample of GRBs that showed simultaneous detections in both the GeV and the X-ray bands during the period from August 2008 to August 2024. To focus on the early-afterglow phase, we considered time intervals starting right after the prompt emission phase and extending up to 10\,ks (see Sect.~\ref{sec:Time-bin selection}). These criteria ensure that the selected intervals correspond to the afterglow phase, where broadband emission is expected to be dominated by external forward shock. We also include hard X-ray (15-150~keV) data observed by BAT to help constrain the potential synchrotron peak. However, because of the limited sensitivity of BAT, the number of such observations is smaller than in only XRT cases. As a result, joint XRT, BAT, and LAT coverage is available only for the early, bright phases of the afterglow emission. The final sample consists of 31 GRBs, with 74 time bins selected for time-resolved spectral analysis after excluding intervals with flaring activity (see Tab.~\ref{tab:timeresolved} and Sect.~\ref{sec:Time-bin selection} for more details). 
 
Using the rich data sample described above, we performed an extensive time-resolved spectral analysis with simultaneous data in X-rays and HE gamma-rays. We compared the fluxes and photon indices, identifying a significant flux-flux correlation (see Fig.~\ref{fig:flux_XRT_LAT}(a)) between X-ray (1~keV) and HE gamma-rays (1~GeV) band. Notably, there is a systematic deviation at lower fluxes, where GeV fluxes become higher relative to X-ray fluxes. However, a correlation with slope close to unity emerges when comparing X-ray emission at 20~keV with 1~GeV (see Fig.~\ref{fig:flux_XRT_LAT}(b)). This suggests that considering only the soft X-ray flux captures only a fraction of the synchrotron energy output, which is lower than the power emitted in the GeV band. The comparison highlights the relative strength power of the emission components at different times. The fluxes at 1 keV and 20 keV represents energy bands where synchrotron emission is the predominant mechanism, while flux at 1 GeV is representative of a synchrotron self-Compton emission. 
Further study using $S_{\gamma}$ (1–1000~keV) as a normalization factor reveals that flux correlations remain stronger for the 20~keV than for the 1~keV (see (a) and (b) of Fig.~\ref{fig:flux_normalized_XRT_LAT}).

In addition, we compared the photon indices in X-ray (0.3--10~keV) with HE gamma-rays. To derive the photon indices in the GeV range, we adopted two approaches: (a) using a fixed energy interval from 0.1–10 GeV, and (b) using a dynamic energy range, defined as 0.1\,GeV–E$_{\mathrm{max}}$, where E$_{\mathrm{max}}$ corresponds to the highest-energy photon detected within a given time bin. This selection of the energy bin is chosen to reduce the systematic uncertainties while calculating the flux in the energy band extending  beyond E\textsubscript{max} (see Fig~\ref{fig:Index_LAT_comparison}). 
By comparing the photon indices in the X-ray and GeV (0.1-10 GeV) energies, we find that indices populate in three broad regimes as seen in the Fig.~\ref{fig:Index_LAT_XRT}: (i) harder X-ray emission having softer HE gamma-ray emission; (ii) moderately harder X-ray emission showing harder HE gamma-ray emission; and (iii) softer X-ray emission. In contrast, there is a shifting of the clustering toward harder GeV photon indices for the dynamic energy range (0.1\,GeV–E\textsubscript{max}) (for details, see Fig~\ref{fig:Index_EmaxLAT_XRTcomparison}). 
The harder X-ray emission with softer or harder HE gamma-ray can correspond to synchrotron and SSC emission, depending on the relative flux levels in the X-ray and HE gamma-ray bands. To confidently identify the dominant radiation process, data from hard X-rays and the MeV range are crucial. In the cases with softer X-ray emission, there is an interesting outlier that may indicate a clear SSC signature.

We explored the possibility of identifying a unique set of microphysical parameters capable of simultaneously explaining X-ray and GeV emission, while reconciling previous independent observational results from X-ray studies (e.g., ~\citealt{2012MNRAS.425..506D}) and GeV studies (e.g., ~\citealt{2014MNRAS.443.3578N}). We have modeled the multiwavelength emission in the afterglow using synchrotron and SSC radiation from a power-law electron distribution accelerated at the forward shock. This approach, described in more detail in Sect.~\ref{sec:Intrepretation}, assumes the self-similar evolution of a relativistic blast wave in a cold circumburst medium to compute broadband spectra at different epochs.

In a simplistic scenario, we assumed that the parameters, such as the fraction of the kinetic energy of the blast-wave distributed to electrons and to the magnetic field, do not change over time. Moreover, we further assumed that the prompt emission efficiency (defined as the ratio of the isotropic equivalent energy in the prompt emission and the kinetic energy of the jet) and the fraction of energy given to the leptons ($\epsilon_{\rm e}$) are both set to the values of 0.1. We found that within the SSC framework, the set of parameters \(p=2.2\), \(\epsilon_{\rm e} = 0.1\), \(\epsilon_{\rm B} = 10^{-4}\), and a wind-medium environment with A$_{*}$ = 0.1, best reproduce the observed trends in both X-ray and GeV emission reported in  ~\citealt{2012MNRAS.425..506D} and ~\citealt{2014MNRAS.443.3578N}, respectively (see Fig.~\ref{fig:LeMoC_1}). Although previous studies \citep[e.g.,][]{2014MNRAS.442.3147B, 2015MNRAS.454.1073B} have reported a preference for a lower values of \(\epsilon_{\rm B}\) ($\sim$ 10$^{-5}$), our results indicate a magnetic energy fraction, $\epsilon_{\rm B}$ $\sim$ 10$^{-4}$, successfully reproduces both X-ray and GeV trends.

Crucially, this configuration also predicts the observed flux-flux (see Fig.~\ref{fig:LeMoC_2}) and index-index (see Fig.~\ref{fig:LeMoC_3}) correlations, supporting the validity of this set of benchmark parameters. In contrast, under the assumptions mentioned above, models that involve a homogeneous medium show systematic deviation from the trends in GeV and X-rays reported in ~\citealt{2014MNRAS.443.3578N}, ~\citealt{2018ApJ...863..138A}, and \citealt{2012MNRAS.425..506D}. However, we note that the flux uncertainties in the GeV band are generally too large to draw firm conclusions based on GeV data alone. Nonetheless, when combining the X-ray and GeV trends together, our analysis shows a preference for a wind-like environment with lower density. 

Although our modeling highlights a set of microphysical parameters that successfully reproduce the observed X-ray and GeV trends, it is important to emphasize that afterglow parameters are inherently highly degenerate. A wide range of combinations involving $\eta$, \(p\), \(\epsilon_{\rm e}\), \(\epsilon_{\rm B}\), and the circumburst density profile can produce broadly similar spectral and temporal behaviors. As a result, the values adopted in this work should be considered as benchmark values that are able to reproduce the main observational trends of population. Our goal is to demonstrate that a physically motivated and self-consistent configuration exists that can account for the observed correlations across multiple energy bands. Future broadband observations particularly in the MeV and TeV regime will be essential for breaking these degeneracies and more tightly constraining the underlying parameter space.

Using the same preferred microphysical parameters, we simulate intrinsic VHE gamma-ray emission (0.3~TeV) in the local Universe. The simulated intrinsic VHE fluxes show a positive correlation with X-ray fluxes (see Fig.~\ref{fig:LeMoC_4}). This correlation is compatible with the trend observed in TeV detected GRBs \citep{MAGIC:2019lau, 2021Sci...372.1081H}. A detailed discussion of the detectability of these VHE emissions, including considerations of instrumental sensitivity, extragalactic background light attenuation, and observational strategies, is beyond the scope of this study and will be discussed in a future work.

We summarize our work as follows:
\begin{itemize}[label=\textbullet, nosep, leftmargin=*]

\item We present a detailed time-resolved spectral analysis of GRBs observed between 2008-2024 jointly detected in X-rays and GeV. With the help of a synchrotron self-Compton model, we attempted to explain the observables (fluxes and photon indices). Extending the set of data and with a new systematic analysis we confirm the correlations found in X-ray and HE gamma-rays. We identify preferred microphysical parameters able to reproduce the observables from this study. We have further estimated expected TeV afterglow emission.

\item From the observations, we identified a flux-flux correlation between X-rays and gamma-rays. However, at lower fluxes, an excess in the GeV power can be identified. A correlation with slope close to unity has been observed between X-rays at 20\,keV and HE gamma-rays (1~GeV).

\item To interpret the observational data, we used a synchrotron self-Compton model and explored the parameter space that influences the afterglow emission. Specifically, we varied the index of the electron energy distribution as well as the fraction of energy allocated to the magnetic field and the medium (wind or homogeneous, with different density values) while keeping the prompt emission efficiency ($\eta$) and the energy fraction in the leptons ($\epsilon_e$) fixed. We simulated observed spectra of 220 long GRBs (similar number of long GRBs observed in a year by Fermi/GBM), assigning isotropic equivalent energy and redshift according to known distributions of the two parameters mentioned above. The afterglow parameters are chosen so that they describe the combined temporal decline of X-ray and GeV fluxes scaled with the isotropic equivalent energy. Our study predicts a set of microphysical parameters, $\eta\sim$0.1, $p\sim$2.2, $\epsilon_e\sim0.1$, $\epsilon_B\sim10^{-4}$, and wind-medium with lower density (A$_{*}\sim 0.1$) is preferred.

\item The preferred parameter set reproduces the flux and index correlations covering both the X-ray and HE gamma-rays. However, we noticed an outlier through index correlation, GRB160325A observed with a softer X-ray spectra (index$<$ -2.2). A detailed study on this GRB will be discussed in a future publication. 

\item The preferred parameters are used to predict the nature of the TeV afterglow. We substantiate the claims from the previously detected GRBs in VHE gamma-rays of the existence of an intrinsic correlation between the keV emission and the intrinsic (EBL unabsorbed) TeV emission. This is crucial to determine the detectability of the GRBs in VHE gamma-rays in the era of the future generation ground based IACTs.

\end{itemize}

\begin{acknowledgements}
The authors acknowledge the use of public data from Fermi and Swift missions and thank the respective teams for making the data available. BB and MB acknowledge financial support from the Italian Ministry of University and Research (MUR) for the PRIN grant METE under contract no. 2020KB33TP.
LN acknowledge financial support from the European Union-Next Generation EU, PRIN 2022 RFF M4C21.1 (202298J7KT - PEACE) and from the INAF grant 'Shock acceleration in Gamma Ray Bursts'. DM acknowledges “funding by the European UnionNextGenerationEU” RFF M4C2 project IR0000012 CTA+.

\end{acknowledgements}

\bibliographystyle{aa}
\bibliography{bibliography}  


\appendix 

\section{Appendix} \label{sec:appendix}

This appendix provides supplementary tables referenced in the main text, presenting detailed data and analysis results for the GRBs studied in this work.

\subsection{GRBs with X-ray flares}
Table~\ref{tab:Flare_GRBs} lists all GRBs in our sample from 2008-2024, that exhibit flaring activity, as identified using the automated \textit{Swift} pipeline \citep{Willingale:2006zh}. The detailed study of these X-ray flares is beyond the scope of the present work.

\begin{table}[ht!]
\centering
\renewcommand{\arraystretch}{1.2}
\setlength{\tabcolsep}{33pt}
\begin{tabular}{|c|c|}
\hline
{GRB} & {Time-bin [s]} \\
\hline
\multirow{2}{*}{100728A} & 154-253 \\
                         & 253-893 \\ \hline
140323A                  & 103-157 \\ \hline
160325A                  & 155-230 \\ \hline
170405A                  & 130-230 \\ \hline
170906A                  & 71-179 \\ \hline
\multirow{3}{*}{171120A} & 229-370 \\
                         & 370-550 \\
                         & 550-583 \\ \hline
181020A                  & 163-454 \\ \hline
210410A                  & 78-111 \\  \hline
\end{tabular}
\caption{The table presents GRBs and its flares identified using the automated Swift pipeline \citep{Willingale:2006zh}. The time bins are given with respect to the BAT trigger time.}
\label{tab:Flare_GRBs}
\end{table}

\subsection{Multiwavelength Spectral Analysis Results}
In this section, we present the analysis results from the XRT, BAT, joint XRT+BAT, and LAT observations. Table~\ref{tab:timeresolved} presents the results of time-resolved spectral analysis for selected GRBs. The time intervals were chosen according to the criteria described in Section~\ref{sec:Time-bin selection}. For most GRBs, the XRT spectral analysis covers the 0.3--10~keV energy range, except for GRB\,160325A, where the range is limited to 1--10~keV due to strong absorption below 1~keV. For each time bin, the table reports the XRT spectral fit parameters, as well as the LAT flux and photon index values for two energy bands: 0.1--10~GeV and 0.1\,GeV--E$_{\rm max}$, where E$_{\rm max}$ is the highest-energy photon detected in each time bin. The table also provides the LAT test statistic (TS) values for both energy bands and indicates whether a BAT observation was recorded during the respective time interval. In addition, we also report flux at 1 keV and 1 GeV.

Table~\ref{tab:nh_stats_unknownz} shows the XRT spectral fitting results for GRBs with unknown redshift, providing results for both $z=0$ and $z=2$ assumptions. Table~\ref{tab:XRT_knownz} summarizes the results of XRT spectral fitting for GRBs with known redshift.The table includes the intrinsic absorption column density ($N_{\rm H}$) and key spectral statistics.
Table~\ref{tab:bat_results} summarizes the BAT results for time bins with available BAT observations. Tables~\ref{tab:xrt_bat_PL} and~\ref{tab:xrt_bat_SBPL} present the joint spectral analysis results for time bins best fit by a simple power law and a smoothly broken power law model, respectively. Each table reports the photon index ($\alpha$) and the flux calculated in the 0.3--150~keV energy range and at 10~keV for the corresponding time bins. Table~\ref{tab:Pl_sbpl} present the comparison between statistic of PL and SBPL for joint XRT and BAT fit.

\begin{table*}[htbp]
\scriptsize

\setlength{\tabcolsep}{2.0pt} 
\begin{tabular}
{| c | c | c c c | c c c c |c| c c c | c |}
\hline
\multirow{4}{*}{GRB}& \multirow{4}{*}{T - T$_{0}^{\rm BAT}$[s] \ } & \multicolumn{3}{c|}{Swift-XRT}            &  \multicolumn{8}{c|}{Fermi-LAT}          & \multirow{4}{*}{BAT} \\  \cline{3-13}
                    &                                        & \multicolumn{2}{c|}{${\rm 0.3-10\,keV}$} & \multicolumn{1}{c|}{${\rm 1\,keV}$} & \multicolumn{3}{c|}{${\rm 0.1-10\,GeV}$} & \multicolumn{1}{c|}{${\rm 1\,GeV}$} & \multirow{2}{*}{E$_{\rm max}$} & \multicolumn{3}{c|}{${\rm 0.1-E_{max} [GeV] }$} & \\ \cline{3-9} \cline{11-13}
                    &     & Flux $\times10^{-10}$ &\multirow{2}{*}{-$\Gamma_{X}$} & Flux $\times10^{-10}$ & Flux $\times10^{-9}$ &  \multirow{2}{*}{-$\Gamma_{\gamma}$} & \multirow{2}{*}{TS} & Flux $\times10^{-9}$& & Flux$\times10^{-9}$&\multirow{2}{*}{$-\Gamma_{\gamma}^{\prime}$} &\multirow{2}{*}{TS} &  \\ 
                    &   & erg cm$^{-2}$ s$^{-1}$ &                       & erg cm$^{-2}$ s$^{-1}$   & erg cm$^{-2}$ s$^{-1}$ &  &  & erg cm$^{-2}$ s$^{-1}$ & [GeV] & erg cm$^{-2}$ s$^{-1}$&  &  & \\ \hline               
081203A & 209-640 & 5.50$\pm$0.09 & 1.70$\pm$0.03 & 0.50$\pm$0.02 & 1.38$\pm$1.08 & 2.12$\pm$0.49 & 24 & 0.28$\pm$0.14 & 1.02 & 1.03$\pm$0.74 & 1.57$\pm$0.70 & 25 & \ding{51}\\
        & 650-798 & 2.56$\pm$0.25 & 1.75$\pm$0.19 & 0.26$\pm$0.08 & $<3.25$ & 2.00 & 0 &  &  &  &  &  & \ding{51}\\ \hline
090510  & 100-262 & 2.34$\pm$0.11 & 1.75$\pm$0.08 & 0.24$\pm$0.03 & 1.81$\pm$1.00 & 3.07$\pm$0.81 & 12 & 0.16$\pm$0.14 & 0.22 & 1.25$\pm$0.67 & -0.01$\pm$0.18 & 13 & \ding{51}\\
        & 262-1503 & 0.88$\pm$0.04 & 1.61$\pm$0.08 & 0.06$\pm$0.01 & $<0.31$ & 2.00 & 0 &  &  &  &  &  & \ding{55}\\
        & 5320-7192 & 0.03$\pm$0.00 & 1.96$\pm$0.06 & 0.00$\pm$0.00 & $<0.25$ & 2.00 & 0 &  &  &  &  &  & \ding{55}\\ \hline
091127A & 5440-5584&2.20$\pm$0.12 & 2.10$\pm$0.11 & 0.44$\pm$0.04 & $<2.50$ & 2.00 & 0 &  &  &  &  &  & \ding{55}\\ \hline
100728A & 893-1325 & 7.37$\pm$0.21 & 1.82$\pm$0.05 & 0.89$\pm$0.06&2.23$\pm$1.68 & 1.23$\pm$0.52 & 12 & 0.28$\pm$0.14 & 1.39 & 1.04$\pm$0.67 & -0.01$\pm$0.03 & 19 & \ding{55}\\
        & 1335-1993 & 5.37$\pm$0.26 & 1.71$\pm$0.10 & 0.5$\pm$0.08&$<1.12$ & 2.00 & 3 &  &  &  &  &  & \ding{55}\\
        & 5388-7458 & 1.24$\pm$0.06 & 1.81$\pm$0.10 & 0.15$\pm$0.2&$<0.36$ & 2.00 & 1 &  &  &  &  &  & \ding{55}\\ \hline
110213A & 1029-1996 & 8.48$\pm$0.31 & 1.80$\pm$0.07 & 0.98$\pm$0.10 & 2.29$\pm$2.18 & 1.41$\pm$0.48 & 20 & 0.35$\pm$0.24 & 3.00 & 1.50$\pm$5.04 & 0.79$\pm$1.00 & 22 & \ding{55}\\
        & 5862-7730 & 3.89$\pm$0.16 & 1.67$\pm$0.08 & 0.33$\pm$0.04 & $<0.99$ & 2.00 & 3 &  &  &  &  &  & \ding{55}\\ \hline
110625A & 146-240 & 39.00$\pm$11.20 & 2.09$\pm$0.23 & 7.50$\pm$2.10 & 29.50$\pm$16.30 & 2.13$\pm$0.41 & 17 & 6.10$\pm$2.10 &  &  &  &  & \ding{51}\\
        & 245-1336 & 6.35$\pm$0.76 & 1.24$\pm$0.24 & 0.15$\pm$0.09 & 8.41$\pm$2.19 & 2.44$\pm$0.25 & 48 & 1.50$\pm$0.30 & 2.51 & 7.73$\pm$1.82 & 2.33$\pm$0.28 & 49 & \ding{55}\\ \hline
110731A & 73-640 & 9.37$\pm$1.17 & 1.96$\pm$0.02 & 1.50$\pm$0.12 & 2.31$\pm$1.87 & 1.28$\pm$0.52 & 25 & 0.30$\pm$0.22 & 3.45 & 1.65$\pm$6.67 & 0.54$\pm$0.87 & 27 & \ding{51}\\
        & 648-1410 & 1.94$\pm$0.08 & 1.84$\pm$0.08 & 0.25$\pm$0.03 & $<0.96$ & 2.00 & 1 &  &  &  &  &  & \ding{55}\\
        & 6053-7189 & 0.26$\pm$0.02 & 1.85$\pm$0.13 & 0.03$\pm$0.01 & $<1.02$ & 2.00 & 1 &  &  &  &  &  & \ding{55}\\ \hline
120729A & 393-635 & 1.78$\pm$0.14 & 1.84$\pm$0.15 & 0.22$\pm$0.05 & 9.32$\pm$7.06 & 1.24$\pm$0.43 & 22 & 1.20$\pm$0.74 & 2.44 & 5.79$\pm$3.43 & 0.48$\pm$0.67 & 26 & \ding{51}\\
        & 4187-6413 & 0.17$\pm$0.09 & 1.93$\pm$0.11 & 0.02$\pm$0.01 & $<0.48$ & 2.00 & 0 &  &  &  &  &  & \ding{55}\\ \hline
121011A & 104-138 & 5.35$\pm$0.44 & 1.65$\pm$0.16 & 0.43$\pm$0.12 & $<21.7$ & 2.00 & 1 &  &  &  &  &  & \ding{51}\\
        & 3475-5875 & 0.02$\pm$0.00 & 1.97$\pm$0.27 & 0.00$\pm$0.00 & $<0.60$ & 2.00 & 1 & & &  &  &  & \ding{55}\\ \hline
130427A & 198-301 & 870.00$\pm$21.00 & 2.05$\pm$0.04 & 160.00$\pm$6.50 & 76.20$\pm$13.4 & 2.29$\pm$0.14 & 505& 15.0$\pm$1.7 & 57.42 & 55.40$\pm$11.30 & 2.21$\pm$0.13 & 526 & \ding{51}\\
        & 301-463 & 317.00$\pm$5.00 & 1.68$\pm$0.03 & 27.00$\pm$1.40 & 49.90$\pm$8.10 & 2.35$\pm$0.14 & 482& 9.7$\pm$1.0 & 4.75 & 47.80$\pm$7.00 & 2.34$\pm$0.13 & 484 & \ding{51}\\
        & 463-500 & 206.00$\pm$6.00 & 1.50$\pm$0.05 & 11.00$\pm$1.00 & 49.80$\pm$23.8 & 1.90$\pm$0.30 & 85 & 10.0$\pm$3.0&7.05 & 47.40$\pm$21.4 & 1.87$\pm$0.30 & 85 & \ding{51}\\ \hline
140102A & 63-885 & 11.90$\pm$0.10 & 1.73$\pm$0.02 & 1.20$\pm$0.03 & $<5.40$ & 2.00 & 35 &  & 8.03 &  &  &  & \ding{51}\\
        & 5652-6633 & 0.35$\pm$0.03 & 1.77$\pm$0.18 & 0.04$\pm$0.01 & $<0.55$ & 2.00 & 1 &  &  &  &  &  & \ding{55}\\ \hline
140323A & 157-214 & 19.5$\pm$2.30 & 2.66$\pm$0.12 & 5.90$\pm$0.44 & $<2.35$ & 2.00 & 15& &  &  & & & \ding{51}\\
        & 214-283 & 4.77$\pm$0.53 & 2.14$\pm$0.16 & 1.00$\pm$0.14 & $<1.39$ & 2.00 & 5& &  &  &  &  & \ding{51}\\
        & 288-696 & 3.41$\pm$0.36 & 2.09$\pm$0.15 & 0.68$\pm$0.09 & $<4.38$ & 2.00 & 7& &  &  &  &  & \ding{51}\\ \hline
150314A & 91-778 & 0.21$\pm$0.01 & 1.75$\pm$0.01 & 0.02$\pm$0.00 & 0.55$\pm$0.26 & 4.44$\pm$2.09 & 12 & 0.00$\pm$0.04 &  &  &  &  & \ding{51}\\
        & 5279-7526 & 0.92$\pm$0.04 & 1.67$\pm$0.07 & 0.08$\pm$0.01 & $<0.34$ & 2.00 & 3 &  &  &  &  &  & \ding{55}\\ \hline
150403A & 254-1787 & 24.80$\pm$0.10 & 1.68$\pm$0.01 & 2.20$\pm$0.03 & 1.54$\pm$0.66 & 1.83$\pm$0.28 & 46& 0.32$\pm$0.09& 5.39 & 1.39$\pm$0.53 & 1.76$\pm$0.31 & 47 & \ding{55}\\
        & 6852-7532 & 3.26$\pm$0.13 & 1.67$\pm$0.07 & 0.28$\pm$0.03 & $<1.38$ & 2.00 & 0 &  & & &  &  & \ding{55}\\ \hline
151006A & 210-587 & 5.15$\pm$0.18 & 1.38$\pm$0.07 & 0.19$\pm$0.03 & 1.63$\pm$0.00 & 2.00 & 1& &  &  &  &  & \ding{51}\\
        & 587-1101 & 2.20$\pm$0.17 & 1.52$\pm$0.18 & 0.12$\pm$0.04 & 1.11$\pm$1.03 & 2.40$\pm$0.74 & 10& 0.18e-10$\pm$0.11 & 0.18$\pm$0.14 &  &  &  & \ding{55}\\ \hline
160325A & 73-155 & 15.91$\pm$0.63 & 2.47$\pm$0.10 & 4.50$\pm$0.16 & 7.64$\pm$4.53 & 2.26$\pm$0.56 & 21& 1.5$\pm$0.55 & 0.99 & 5.67$\pm$2.76 & 1.46$\pm$0.94 & 23 & \ding{51}\\
        & 230-504 & 4.41$\pm$0.18 & 2.50$\pm$0.10 & 1.3$\pm$0.05 &1.50$\pm$0.81 & 2.83$\pm$0.68 & 20 & 0.17$\pm$0.01 &0.53 & 1.55$\pm$0.64 & 2.23$\pm$0.87 & 24 & \ding{51}\\ \hline
160905A & 65-1100 & 28.00$\pm$0.30 & 1.81$\pm$0.02 & 3.30$\pm$0.09 & 3.40$\pm$1.30 & 1.64$\pm$0.23 & 70 & 0.65$\pm$0.16 & 7.94 & 3.21$\pm$1.18 & 1.61$\pm$0.23 & 70 & \ding{51}\\ 
        & 5500-6000 & 1.87$\pm$0.34 & 2.12$\pm$0.23 & 0.38$\pm$0.08 & $<3.22$ & 2.00 & 2 &  & & &  &  & \ding{55}\\ \hline
160917A & 101-136 & 6.92$\pm$0.61 & 1.26$\pm$0.18 & 0.17$\pm$0.05&$<17.1$ & 2.00 & 0 &  &  &  &  &  & \ding{51}\\ \hline
170405A & 230-600 & 6.10$\pm$0.12 & 1.80$\pm$0.02 & 0.71$\pm$0.2 &2.07$\pm$0.79 & 1.99$\pm$0.37 & 30& 0.43$\pm$0.1 & 0.89 & 2.07$\pm$0.79 & 1.07$\pm$0.62 & 35 & \ding{51}\\
        & 620-880 & 1.99$\pm$0.12 & 1.72$\pm$0.11 & 0.19$\pm$0.03 & 1.77$\pm$1.21 & 2.25$\pm$0.54 & 18 & 0.33$\pm$0.16 & 0.46 & 1.32$\pm$0.61 & 0.29$\pm$1.12 & 25 & \ding{51}\\
        & 4600-6500 & 0.24$\pm$0.11 & 2.03$\pm$0.09 & 0.04$\pm$0.01 & $<0.43$ & 2.00 & 3 &  &  &  &  &  & \ding{55}\\ \hline
170728B & 474-697 & 4.81$\pm$0.20 & 2.07$\pm$0.07 & 0.93$\pm$0.06 & $<2.03$ & 2.00 & 0 & & &  &  &  & \ding{51}\\
        & 703-1608 & 3.28$\pm$0.16 & 1.88$\pm$0.09 & 0.45$\pm$0.05 & $<1.47$ & 2.00 & 1 & & &  &  &  & \ding{55}\\
        & 6231-7461 & 0.58$\pm$0.04 & 1.98$\pm$0.12 & 0.10$\pm$0.01 & $<0.49$ & 2.00 & 1 & & &  &  &  & \ding{55}\\ \hline
170813A & 142-160 & 2.82$\pm$0.54 & 1.44$\pm$0.35& 0.12$\pm$0.06 & $<28.7$ & 2.00 & 0 &  & & &  &  & \ding{51}\\
        & 160-340 & 1.55$\pm$0.22 & 1.64$\pm$0.32& 0.12$\pm$0.05 & $<5.13$ & 2.00 & 0 &  & & &  &  & \ding{51}\\ \hline
170906A & 179-286 & 18.47$\pm$0.59 & 2.01$\pm$0.05 & 3.2$\pm$0.17 &22.20$\pm$8.60 & 1.39$\pm$0.37 & 49 & 3.4$\pm$0.93 & 3.60 & 16.9$\pm$8.60 & 1.06$\pm$0.48 & 51 & \ding{51}\\
        & 286-750 & 13.06$\pm$0.21 & 1.80$\pm$0.03 & 1.5$\pm$0.06&5.29$\pm$2.01 & 2.07$\pm$0.28 & 63 & 1.1$\pm$0.27 & 2.55 & 4.40$\pm$1.38 & 1.88$\pm$0.34 & 64 & \ding{55}\\ \hline
171120A & 104-229 & 13.82$\pm$2.23 & 2.06$\pm$0.16 & 2.60$\pm$0.46 & $<6.17$ & 2.00 & 2 &  &  &  &  &  & \ding{51}\\
        & 4400-6400 & 0.32$\pm$0.06 & 1.92$\pm$0.23 & 0.05$\pm$0.01 &0.50$\pm$0.33 & 1.76$\pm$0.45 & 18 & 0.10$\pm$0.04 & 3.43 & 0.39$\pm$0.22 & 1.52$\pm$0.56 & 19 & \ding{55}\\ \hline
180720B & 109-147 & 588.0$\pm$12.00 & 1.77$\pm$0.04 & 64.00$\pm$3.70 & 69.60$\pm$30.80 & 1.94$\pm$0.29 & 80 & 15.00$\pm$4.10 & 4.91 & 62.20$\pm$24.20 & 1.87$\pm$0.31 & 80 & \ding{51}\\
        & 147-625 & 86.70$\pm$5.00 & 1.73$\pm$0.01 & 8.50$\pm$0.32 & 14.60$\pm$3.20 & 2.09$\pm$0.16 & 212 & 3.10$\pm$0.42 & 1.28 & 14.60$\pm$3.20 & 1.67$\pm$0.21 & 223 & \ding{51}\\ \hline
181020A & 454-1080 & 7.02$\pm$0.08 & 1.95$\pm$0.02 & 1.10$\pm$0.03 & $<1.44$ & 2.00 & 0 &  & && &  & \ding{55}\\
        & 4850-6800 & 0.55$\pm$0.03 & 2.01$\pm$0.10 & 0.10$\pm$0.01 & $<0.94$ & 2.00 & 4 &  & && &  & \ding{55}\\ \hline
190511A & 130-300 & 5.23$\pm$2.01 & 2.07$\pm$0.07 & 1.00$\pm$0.24 & 2.87$\pm$1.69 & 2.60$\pm$0.61 & 15 & 0.42$\pm$0.25 & 0.49&2.45$\pm$1.10&1.37$\pm$0.92 & 16 & \ding{55}\\
        & 300-381 & 2.94$\pm$0.37 & 1.97$\pm$0.25 & 0.48$\pm$0.13 & $<5.73$ & 2.00 & 0 &  & && &  & \ding{55}\\
        & 5697-6119 & 0.13$\pm$0.03 & 2.21$\pm$0.34 & 0.03$\pm$0.01 & $<19.10$ & 2.00 & 0 &  & && &  & \ding{55}\\ \hline
200716C & 87-118 & 69.90$\pm$1.62 & 1.35$\pm$0.05 & 2.30$\pm$0.24 & $<19.10$ & 2.00 & 0 &  & && &  & \ding{51}\\
        & 118-148 & 52.19$\pm$1.34 & 1.39$\pm$0.05 & 2.00$\pm$0.20 & $<22.50$ & 2.00 & 0 &  & && &  & \ding{51}\\
        & 148-342 & 16.15$\pm$0.26 & 1.78$\pm$0.03 & 1.80$\pm$0.08 & 7.03$\pm$4.71 & 1.7$\pm$0.42 & 19 & 1.40$\pm$0.60 & 8.17 & 6.72$\pm$4.34 & 1.68$\pm$0.43 & 19 & \ding{55}\\
        & 342-670 & 3.50$\pm$0.09 & 1.66$\pm$0.05 & 0.29$\pm$0.02 & $<4.73$ & 2.00 & 1 & &&& &  & \ding{55}\\
        & 670-1050 & 1.88$\pm$0.12 & 1.53$\pm$0.11 & 0.11$\pm$0.02 & $<3.33$ & 2.00 & 0 & &&& &  & \ding{55}\\
        & 4284-5048 & 0.33$\pm$0.02 & 1.35$\pm$0.12 & 0.01$\pm$0.00 & $<7.09$ & 2.00 & 5 & &&& &  & \ding{55}\\ \hline
210410A & 111-186 & 2.47$\pm$0.18 & 1.70$\pm$0.16 & 0.23$\pm$0.06 & $<18.70$ & 2.00 & 4 & &&& &  & \ding{55}\\ 
        & 195-1091 & 0.66$\pm$0.04 & 1.39$\pm$0.12 & 0.02$\pm$0.01 & 5.08$\pm$3.06 & 2.21$\pm$0.49 & 18& 1.00$\pm$0.37 & 8.17 & 3.94$\pm$1.76 & 1.54$\pm$0.69 & 20 & \ding{55}\\
        & 4342-5000 & 0.07$\pm$0.01 & 1.99$\pm$0.34 & 0.01$\pm$0.00 & $<1.02$ & 2.00 & 1 & &&& &  & \ding{55}\\ \hline
210619B & 334-1864 & 22.13$\pm$0.01 & 1.90$\pm$0.01 & 3.20$\pm$0.04 & 2.29$\pm$1.27 & 1.10$\pm$0.38 & 31& 0.25$\pm$0.09 & 8.38 & 2.16$\pm$1.17 & 1.03$\pm$0.41 & 32 & \ding{51}\\
        & 6400-7565 & 3.37$\pm$0.11 & 1.92$\pm$0.06 & 0.50$\pm$0.04 & $<0.60$ & 2.00 & 1 & &&& &  & \ding{55}\\ \hline
220101A & 3807-5143 & 3.86$\pm$0.16 & 1.63$\pm$0.07 & 0.29$\pm$0.03 & $<1.26$ & 2.00 & 4& & $<0.1$&& &  & \ding{55}\\
        & 9362-10000 & 1.39$\pm$0.07 & 1.59$\pm$0.09 & 0.09$\pm$0.02 & $<53.10$ & 2.00 & 0& & && &  & \ding{55}\\ \hline
240825A & 417-1294 & 31.07$\pm$0.23 & 1.70$\pm$0.02 & 2.80$\pm$0.09 & 1.09$\pm$0.63 & 2.12$\pm$0.42 & 22& 0.22$\pm$0.08& 1.38 & 0.84$\pm$0.37 & 1.75$\pm$0.54 & 23 & \ding{51}\\
        & 4796-6378 & 1.91$\pm$0.10 & 1.59$\pm$0.11 & 0.13$\pm$0.03 & $<76.20$ & 2.00 & 0 & &&& &  & \ding{55}\\ \hline
\end{tabular}
\caption{Time-resolved spectral analysis results for GRBs with time intervals selected according to the criteria described in Sect.~\ref{sec:Time-bin selection}. The XRT columns present the results of spectral analysis for the 0.3–10~keV energy range, except for GRB160325A$^{\star}$ where the range is 1–10~keV due to strong absorption below 1\,keV. XRT column also report the flux at 1~keV. The LAT columns report flux and photon index values for two energy ranges: 0.1–10~GeV and 0.1–E$_{\rm max}$, where E$_{\rm max}$ is the highest-energy photon detected within each time bin. The flux and upper limit at 1~GeV and the test statistic (TS) values corresponding to the LAT detections are also provided for both energy bands. The final column indicates whether a BAT observation was recorded during the respective time interval. Flux values are given in units of erg cm$^{-2}$ s$^{-1}$, and photon indices ($\Gamma$) characterize the spectral slopes in given energy band.}
\label{tab:timeresolved} 
\end{table*}

\begin{table*}[ht]
    \centering
    \small
    \renewcommand{\arraystretch}{1.2}
    \setlength{\tabcolsep}{2.5pt}
    \begin{tabular}{|c|c|c|c|c|c|c|c|c|c|}
        \hline
\multirow{4}{*}{GRB}     & \multirow{3}{*}{T - T$_{0}^{\rm BAT}$ \ }     & \multicolumn{4}{c|}{z=0}                                                                          & \multicolumn{4}{c|}{z=2} \\ \cline{3-10}
                         &                               & Flux                   & \multirow{3}{*}{-$\Gamma_{X}$}   & \multirow{3}{*}{cstat/d.o.f} & $N_{\rm H}$           & Flux             & \multirow{3}{*}{-$\Gamma_{X}$}  & \multirow{3}{*}{cstat/d.o.f} & $N_{\rm H}$   \\ 
                         & [s]                           & $\times10^{-10}$       &                                  &               & (Intrinsic)           & $\times10^{-10}$ &                                 &               & (Intrinsic)              \\ 
                         &                               & erg cm$^{-2}$ s$^{-1}$ &                                  &               & [$10^{22}$ cm$^{-2}$] & erg cm$^{-2}$ s$^{-1}$ &                           &               & [$10^{22}$ cm$^{-2}$]     \\ \hline
\multirow{2}{*}{110625A} & 146-240            & 39.00$\pm$11.20& 2.09$\pm$0.23      & 375/467 & 5.04$\pm$0.74   & 45.00$\pm$19.63  &2.09$\pm$0.32 & 409/467 & 82.01$\pm$15.91 \\ 
                         & 245-1336           & 6.35$\pm$0.76& 1.24$\pm$0.24        & 309/356 & 3.39$\pm$0.72   & 7.94$\pm$1.96    &1.42$\pm$0.33 & 320/356 & 63.10$\pm$16.61\\ \hline
                         
\multirow{2}{*}{121011A} & 104-138            & 5.35$\pm$0.44& 1.65$\pm$0.16        & 163/378 & 0.05$\pm$0.04   & 5.26$\pm$0.42    &1.61$\pm$0.35 & 162/378 & $<$1.01            \\ 
                         & 3475-5875          & 0.02$\pm$0.00& 1.97$\pm$0.27        & 77/85   & 0.01$\pm$0.04   & 0.02$\pm$0.00    & 1.99$\pm$0.22& 76/85   & $<$0.78  \\ \hline
                         
\multirow{2}{*}{140102A} & 63-885             & 11.90$\pm$0.01& 1.73$\pm$0.02       & 724/752 & 0.08$\pm$0.01   & 11.48$\pm$0.01   &1.66$\pm$0.02 & 722/752 & 0.65$\pm$0.05  \\ 
                         & 5652-6633          & 0.35$\pm$0.03& 1.77$\pm$0.18        & 133/400 & 0.05$\pm$0.04   & 0.34$\pm$0.03    &1.76$\pm$0.15 & 132/400 & $<$1.05            \\ \hline
                         
\multirow{3}{*}{140323A} & 157-214            & 19.50$\pm$2.30& 2.66$\pm$0.12       & 311/538 & 0.38$\pm$0.04   & 14.90$\pm$1.17   & 2.46$\pm$0.10 & 310/538 & 3.98$\pm$0.49             \\ 
                         & 214-283            & 4.77$\pm$0.53& 2.14$\pm$0.16        & 198/289 & 0.27$\pm$0.06   & 4.21$\pm$0.32    & 1.99$\pm$0.13 & 195/289 & 2.72$\pm$0.65             \\ 
                         & 288-696            & 3.41$\pm$0.36& 2.09$\pm$0.15        & 216/269 & 0.41$\pm$0.07   & 3.00$\pm$0.22    & 1.92$\pm$0.13 & 214/269 & 4.65$\pm$0.88            \\ \hline
                         
\multirow{2}{*}{151006A} & 210-587            & 5.15$\pm$0.18& 1.38$\pm$0.07        & 508/553 & 0.23$\pm$0.04   & 5.08$\pm$0.18    & 1.29$\pm$0.07 & 510/553 & 2.68$\pm$0.54  \\ 
                         & 587-1101           & 2.20$\pm$0.17& 1.52$\pm$0.18        & 168/189 & 0.22$\pm$0.08   & 2.16$\pm$0.16    & 1.42$\pm$0.15 & 167/189 & 2.38$\pm$0.93      \\ \hline

\multirow{2}{*}{160325A$^{\star}$} & 73-155   & 15.91$\pm$0.63& 2.47$\pm$0.10        & 445/474 & $<$0.02           & 15.90$\pm$0.62   & 2.46$\pm$0.08 & 445/474 & $<$0.34 \\ 
                         & 230-504            & 4.41$\pm$0.18& 2.50$\pm$0.10        & 422/449 & $<$0.02           & 4.41$\pm$0.12     & 2.50$\pm$0.05 & 422/449 & $<$0.45  \\ \hline

\multirow{2}{*}{160905A} & 65-1100           & 28.00$\pm$0.30& 1.81$\pm$0.02        & 908/851 & 0.04$\pm$0.01   & 27.76$\pm$0.22  & 1.80$\pm$0.01     & 904/851 & 0.45$\pm$0.09\\ 
                         & 5500-6000         & 1.87$\pm$0.34& 2.12$\pm$0.23         & 127/269 & 0.25$\pm$0.13   & 1.78$\pm$0.27   & 2.05$\pm$0.21     & 128/269 & $<$8.90 \\ \hline
                         
160917A                  & 101-136           & 6.92$\pm$0.61&1.26$\pm$0.18          & 176/379 & 0.04$\pm$0.08   &6.89$\pm$0.60 & 1.26$\pm$0.15 & 190/379 &  $<$1.98 \\ \hline

\multirow{2}{*}{170813A} & 142-160           & 2.82$\pm$0.54& 1.44$\pm$0.35         & 64/332 & $<$0.22              & 2.79$\pm$0.54& 1.39$\pm$0.29         & 64/332 & $<$1.66     \\ 
                         & 160-340           & 1.55$\pm$0.22& 1.64$\pm$0.32         & 66/298 & 0.19$\pm$0.11      & 1.48$\pm$0.19& 1.54$\pm$0.27         & 65/298 & 1.85$\pm$1.17     \\ \hline
                         
\multirow{2}{*}{170906A} & 179-286           & 18.47$\pm$0.59& 2.01$\pm$0.05        & 517/622 & 0.08$\pm$0.02   & 18.15$\pm$0.45 &2.01$\pm$0.04   & 493/622 & 0.91$\pm$0.15\\
                         & 286-750           & 13.06$\pm$0.21& 1.80$\pm$0.03        & 622/697 & 0.11$\pm$0.01   & 12.59$\pm$0.17 &1.74$\pm$0.03   & 604/697 & 1.11$\pm$0.11\\ \hline
                         
\multirow{2}{*}{171120A} & 104-229            &13.82$\pm$2.23& 2.06$\pm$0.16         & 366/448 & 2.00$\pm$0.21   &14.85$\pm$2.40& 2.01$\pm$0.16   & 366/448 & 37.05$\pm$4.29\\ 
                         & 4400-6400          &0.32$\pm$0.06  & 1.92$\pm$0.23   & 195/226 & 1.87$\pm$0.31         &0.32$\pm$0.06& 1.80$\pm$0.23   & 199/226 & 31.75$\pm$5.96\\ \hline 

\multirow{3}{*}{190511A} & 130-300            & 5.23$\pm$2.01 & 2.07$\pm$0.07   & 369/463 & $<$0.02 & 5.25$\pm$1.20 & 2.07$\pm$0.06   & 369/463 & $<$0.19 \\
                         & 300-381            & 2.94$\pm$0.37 & 1.97$\pm$0.25   & 94/206  & $<$0.08 & 2.88$\pm$0.26 & 1.93$\pm$0.20   & 94/206  & $<$1.24\\
                         & 5697-6119          & 0.13$\pm$0.03 & 2.21$\pm$0.34   & 50/150  & $<$0.16 & 0.13$\pm$0.08 & 2.27$\pm$0.28   & 50/150  & $<$1.49\\\hline   
       
\multirow{6}{*}{200716C}& 87-118             & 69.90$\pm$1.62& 1.35$\pm$0.05 & 633/648& 0.09$\pm$0.01 & 69.18$\pm$1.59& 1.26$\pm$0.37 & 640/648& 0.65$\pm$0.11  \\
                        & 118-148            & 52.19$\pm$1.34& 1.39$\pm$0.05 & 561/594& 0.07$\pm$0.01 & 51.63$\pm$1.37& 1.33$\pm$0.04 & 560/594& 0.57$\pm$0.11  \\
                        & 148-342            & 16.15$\pm$0.26& 1.78$\pm$0.03 & 585/634& 0.06$\pm$0.01 & 15.66$\pm$0.24& 1.70$\pm$0.03 & 595/634& 0.45$\pm$0.06  \\
                        & 342-670            & 3.50$\pm$0.09 & 1.66$\pm$0.05 & 448/570& $<$0.03         & 3.48$\pm$0.09 & 1.64$\pm$0.04 & 449/570& $<$0.25          \\
                        & 670-1050           & 1.88$\pm$0.12 & 1.53$\pm$0.11 & 219/277& $<$0.03         & 1.88$\pm$0.12 & 1.53$\pm$0.01 & 219/277& $<$0.29          \\
                        & 4284-5048          & 0.33$\pm$0.02 & 1.35$\pm$0.12 & 193/417& $<$0.05         & 0.33$\pm$0.02 & 1.36$\pm$0.10 & 192/417& $<$0.32          \\ \hline

\multirow{3}{*}{210410A}& 111-186            & 2.47$\pm$0.18 & 1.70$\pm$0.16 & 153/236 & 0.08$\pm$0.04 & 2.40$\pm$0.17 & 1.63$\pm$0.13 & 153/236 & 0.63$\pm$0.35 \\
                        & 195-1091           & 0.66$\pm$0.04 & 1.39$\pm$0.12 & 144/264 & $<$0.06         & 0.66$\pm$0.05 & 1.36$\pm$0.10 & 144/264 & $<$0.39         \\
                        & 4342-5000          & 0.07$\pm$0.01 & 1.99$\pm$0.34 & 57/201 & $<$0.27          & 0.06$\pm$0.01 & 1.91$\pm$0.28 & 57/201 & $<$2.34          \\
\hline
\end{tabular}
\caption{Swift/XRT spectral fitting results for GRBs with unknown redshift. Columns present fluxes (0.3--10~keV), photon indices ($\Gamma_X$), fit statistics (C-stat/d.o.f), and intrinsic absorption column densities ($N_{\rm H}$) for two redshift assumptions: $z=0$ (left) and $z=2$ (right). Note: Analysis for GRB160325A was performed in the 1--10~keV band due to high absorption below 1~keV.}
\label{tab:nh_stats_unknownz}
\end{table*}

\begin{table}[H]
    \centering
    \renewcommand{\arraystretch}{1.412}
    \setlength{\tabcolsep}{7.3pt}
    \begin{tabular}{|c|c|c|c|}
        \hline
\multirow{3}{*}{GRB}     & \multirow{2}{*}{T - T$_{0}^{\rm BAT}$ \ } & \multicolumn{2}{c|}{Swift-XRT}       \\ \cline{3-4}
                         &                           & \multirow{2}{*}{cstat/d.o.f}  & $N_{\rm H}$ (Intrinsic) \\
                         & [s]                       &                               & [$10^{22}$ cm$^{-2}$] \\ 
\hline
\multirow{2}{*}{081203A} & 209-640             & 586/632 & 0.22$\pm$0.06 \\
                         & 650-798             & 97/311  & 0.53$\pm$0.42 \\ \hline
                         
090510                   & 100-262             & 304/332 & 0.15$\pm$0.06 \\ 
                         & 262-1503            & 295/548 & 0.01$\pm$0.05 \\ 
                         & 1503-7192           & 64/242  & 0.22$\pm$0.16 \\ \hline
                         
\multirow{2}{*}{100728A} & 893-1325            & 581/582 & 2.16$\pm$0.25 \\
                         & 1335-1993           & 289/503 & 1.72$\pm$0.04 \\ \hline

\multirow{2}{*}{110213A} & 1029-1996           & 343/391 & 1.35$\pm$0.18 \\
                         & 5862-7730           & 303/352  & 1.42$\pm$0.23 \\ \hline
                         
\multirow{3}{*}{110731A} & 73-640              & 715/731 & 0.14$\pm$0.12 \\ 
                         & 648-1410            & 272/426 & 0.01$\pm$0.40 \\ 
                         & 6053-7189           & 173/334 & 0.54$\pm$0.75 \\ \hline
                         
\multirow{2}{*}{120729A} & 393-635             & 155/264 & 0.10$\pm$0.15 \\
                         & 4187-6413           & 220/489 & 0.08$\pm$0.11 \\ \hline
                         
\multirow{3}{*}{130427A} & 198-301             & 682/643 & 0.07$\pm$0.01 \\
                         & 301-463             & 857/754 & $$<$$0.01 \\
                         & 463-500             & 522/590 & $$<$$0.01 \\ \hline
          
\multirow{2}{*}{150314A} & 91-778              & 931/808 & 0.87$\pm$0.04 \\ 
                         & 5279-7526           & 358/423 & 1.07$\pm$0.21 \\ \hline
                         
\multirow{3}{*}{170405A} & 230-600             & 302/325 & $$<$$0.13 \\
                         & 620-880             & 207/356 & $$<$$0.78 \\ 
                         & 4600-6500           & 297/433 & $$<$$0.78 \\ \hline
                         
\multirow{3}{*}{170728B} & 474-697             & 472/534 & 0.17$\pm$0.02 \\ 
                         & 703-1608            & 377/531 & 0.15$\pm$0.02 \\ 
                         & 6231-7461           & 272/423 & 0.23$\pm$0.04 \\ \hline
                         
\multirow{2}{*}{180720B} & 109-147             & 527/636 & 0.22$\pm$0.08 \\
                         & 147-625             & 958/849 & 0.18$\pm$0.02 \\ \hline

\multirow{2}{*}{181020A} & 454-1080           & 658/654 & 0.52$\pm$0.08\\
                         & 4850-6800          & 212/485& 0.70$\pm$0.33\\ \hline
       
\multirow{2}{*}{210619B}& 334-1864           & 919/866 & 0.24$\pm$0.03 \\
                        & 6400-7565          & 377/589 & 0.52$\pm$0.21 \\\hline

\multirow{3}{*}{220101A}& 100-239            & 883/775 & $$<$$0.05\\
                        & 3807-5143          & 292/352 & $$<$$1.47 \\
                        & 9362-10000         & 244/285 & $$<$$0.94 \\ \hline

\multirow{3}{*}{240825A}& 417-1294          & 865/846 & 0.77$\pm$0.02\\
                        & 4796-6378         & 270/463 & 0.58$\pm$0.12 \\ \hline
\end{tabular}
\caption{This table present X-ray spectral fitting statistics (C-stat/d.o.f) and intrinsic absorption column density ($N_{\rm H}$) for the GRBs with known redshift. Time bins are measured relative to the BAT trigger time, \(T_0^{\rm BAT}\).}
\label{tab:XRT_knownz}
\end{table}


\begin{table*}[ht!]
    \centering
    \setlength{\tabcolsep}{24pt}
    \begin{tabular}{|c|c|c|c|c|} \hline
\multirow{4}{*}{GRB} & \multirow{2}{*}{T - T$_{0}^{\rm BAT}$ \ }  & \multicolumn{3}{c|}{BAT (15.0-150.0 keV)}                     \\ \cline{3-5}
                     &                                            &   Flux                     & \multirow{3}{*}{-$\Gamma_{B}$} & \multirow{3}{*}{stat/d.o.f}     \\ 
                     & \multirow{2}{*}{[s]}                       &   $\times10^{-9}$          &                 &                                        \\ 
                     &                                            &   erg cm$^{-2}$ s$^{-1}$   &                 &                                      \\ \hline
\multirow{2}{*}{081203A} & 209-640 & $<1.08$  & 2.00          & 43/46   \\
                         & 650-798 & $<1.51$  & 2.00          & 36/46   \\ \hline

090510                  & 100-262 &  $<1.16$  & 2.00          & 33/46    \\ \hline

110625A                  & 146-240 &  $<3.44$ & 2.00          & 61/46    \\ \hline

110731A                  & 73-640  &  $<5.65$ & 2.00          & 45/46    \\ \hline

120729A                  & 393-635 &  $<1.04$ & 2.00          & 40/46    \\ \hline

121011A                  & 104-138 &  $<2.45$ & 2.00          & 51/46    \\ \hline

\multirow{3}{*}{130427A} & 198-301  & 63.10$\pm$1.30 & 1.86$\pm$0.05 & 47/45  \\
                         & 301-463 & 37.90$\pm$9.60 & 1.88$\pm$0.06 & 35/45  \\
                         & 463-500 & 28.70$\pm$1.70 & 1.85$\pm$0.17 & 42/45  \\ \hline
                         
140102A                  & 63-885 &  0.91$\pm$0.57  & 2.32$\pm$0.66 & 56/45    \\ \hline

\multirow{3}{*}{140323A} & 157-214 & $<2.10$  & 2.00          & 49/46  \\
                         & 214-283 & $<1.51$  & 2.00          & 32/46  \\ 
                         & 283-697 & $<1.03$  & 2.00          & 48/46  \\ \hline

150314A                  & 92-778 &  $<1.38$ & 2.00          & 42/46    \\ \hline

151006A                  & 210-587 & $<1.73$ & 2.00          & 50/46    \\ \hline

\multirow{2}{*}{160325A} & 73-155  & $<2.06$ & 2.00          & 34/46  \\
                         & 230-504 & $<0.92$ & 2.00         & 44/46  \\  \hline
                         
160905A                  & 65-1100 & $<1.52$ & 2.00          & 36/46    \\ \hline

160917A                  & 102-137 & $<2.04$ & 2.00          & 52/46    \\ \hline

\multirow{2}{*}{170405A} & 230-600 & $<1.26$ & 2.00          & 63/46 \\
                         & 620-880 & $<1.64$ & 2.00          & 416/46 \\ \hline

170728B                  & 474-697 & $<2.49$ & 2.00          & 74/46    \\ \hline

170813A                  & 142-160 & $<5.45$  & 2.00          & 56/46    \\ 
                         & 160-340 & $<1.13$  & 2.00          & 42/46    \\\hline

\multirow{1}{*}{170906A} & 179-286 & $<1.66$  & 2.00          & 42/46    \\   \hline

\multirow{1}{*}{171120A} & 104-229 & $<1.01$ & 2.00          & 48/46  \\ \hline
                        
\multirow{2}{*}{180720B} & 109-147 & 413.00$\pm$18.00 & 1.78$\pm$0.13 & 41/45  \\
                         & 147-625 & 6.69$\pm$0.70 & 2.33$\pm$0.26 & 42/45   \\ \hline

\multirow{2}{*}{200716C} & 87-118  & 6.77$\pm$1.75 & 1.38$\pm$0.77 & 31/45  \\
                         & 118-148 & $<5.38$ & 2.00 & 39/46   \\ \hline

\multirow{1}{*}{210619B} & 334-1864 & 1.48$\pm$0.50  & 1.89$\pm$0.99  & 41/45    \\   \hline

\multirow{1}{*}{220101A} & 100-239  & 76.98$\pm$1.12  & 1.24$\pm$0.04 & 40/45    \\   \hline

\multirow{1}{*}{240825A} & 417-1294 & 1.53$\pm$0.97  & 2.81$\pm$1.33 & 35/45    \\   \hline
    \end{tabular}%
    \caption{GRB spectral fitting results for BAT in the specified time bins relative to the BAT trigger time \(T_0^{\rm BAT}\). The flux (\(C_{\rm flux}\)) is given in units of erg cm\(^{-2}\) s\(^{-1}\) in the 15.0–150.0 keV energy range. The photon index \(\Gamma_{B}\) and the fitting statistic per degrees of freedom (stat/d.o.f) are also listed.}

    \label{tab:bat_results}
\end{table*}

\begin{table}[ht!]
    \centering
    \small 
    \renewcommand{\arraystretch}{1.5}
    \setlength{\tabcolsep}{1.7pt} 
    \begin{tabular}{|c|c|c|c|c|} \hline
\multirow{4}{*}{GRB}     &                 & \multicolumn{3}{c|}{XRT+BAT; 0.3-150\,keV} \\ \cline{3-5}
                         &   T - T$_{0}^{\rm BAT}$ & \multirow{3}{*}{-\(\Gamma_{XB}\)} & F$_{0.3-150\,keV}$ & F$_{20\,keV}$  \\ 
                         &    [s]  &                  & $\times10^{-9}$     & $\times10^{-9}$   \\
                         &         &                  & [erg\,cm$^{-2}$\,s$^{-1}$]   &[erg\,cm$^{-2}$\,s$^{-1}$] \\  \hline
\multirow{2}{*}{081203A} & 209-640 &  1.69$\pm$0.03   & 1.64$\pm$0.11   & 0.32$\pm$0.01 \\
                         & 650-798 &  1.67$\pm$0.15   & 0.81$\pm$0.28   & 0.16$\pm$0.03  \\ \hline
090510                   & 100-262 &  1.72$\pm$0.08   & 0.65$\pm$0.12   & 0.13$\pm$0.01  \\ \hline
110625A                  & 146-240 &  1.87$\pm$0.10   & 6.46$\pm$0.44   & 1.3$\pm$0.05 \\ \hline
110731A                  & 73-640  &  1.98$\pm$0.02   & 1.70$\pm$0.05   & 0.28$\pm$0.01  \\ \hline
120729A                  & 393-635 &  1.85$\pm$0.14   & 0.39$\pm$0.09   & 0.07$\pm$0.01 \\ \hline
121011A                  & 104-138 &  1.57$\pm$0.13   & 2.08$\pm$0.64   & 0.41$\pm$0.06\\ \hline
\multirow{3}{*}{130427A} & 198-301 &  1.95$\pm$0.01   & 161.00$\pm$2.00 & 27.80$\pm$0.20\\
                         & 301-463 &  1.79$\pm$0.01   & 73.4$\pm$0.80   & 14.00$\pm$0.08\\
                         & 463-500 &  1.72$\pm$0.01   & 52.50$\pm$1.30  & 10.00$\pm$0.13\\ \hline
\multirow{3}{*}{140323A} & 157-214 &  2.66$\pm$0.12   & 2.11$\pm$0.18   & 0.07$\pm$0.02\\
                         & 214-283 &  2.11$\pm$0.16   & 0.72$\pm$0.08   & 0.12$\pm$0.01\\
                         & 283-697 &  2.09$\pm$0.14   & 0.54$\pm$0.05   & 0.08$\pm$0.01\\ \hline
151006A                  & 210-587 &  1.44$\pm$0.05   & 2.65$\pm$0.30   & 0.53$\pm$0.02 \\ \hline
\multirow{2}{*}{160325A*}& 73-155  &  2.49$\pm$0.09   & 2.15$\pm$0.10   & 0.14$\pm$0.02\\
                         & 230-504 &  2.52$\pm$0.09   & 0.59$\pm$0.03   & 0.04$\pm$0.01\\ \hline
160905A                  & 65-1110 &  1.89$\pm$0.01   & 5.96$\pm$0.11   & 1.00$\pm$0.01\\ \hline
160917A                  & 102-137 &  1.50$\pm$0.14   & 2.87$\pm$0.90   & 0.57$\pm$0.07 \\ \hline
\multirow{2}{*}{170405A} & 230-600 &  1.67$\pm$0.03   & 1.91$\pm$0.14   & 0.37$\pm$0.01\\
                         & 620-880 &  1.67$\pm$0.11   & 0.56$\pm$0.14   & 0.11$\pm$0.01\\ \hline
170728B                  & 474-697 &  1.99$\pm$0.06   & 0.79$\pm$0.06   & 0.13$\pm$0.01\\ \hline
\multirow{2}{*}{170813A} & 142-160 &  1.52$\pm$0.27   & 1.12$\pm$0.77   & 0.21$\pm$0.06\\
                         & 160-340 &  1.71$\pm$0.28   & 0.50$\pm$0.25   & 0.09$\pm$0.02 \\ \hline
\multirow{1}{*}{170906A} & 179-286 &  2.01$\pm$0.04   & 3.17$\pm$0.16   & 0.51$\pm$0.02\\ \hline
\multirow{1}{*}{171120A} & 104-229 &  2.25$\pm$0.16   & 2.26$\pm$0.14   & 0.30$\pm$0.02 \\ \hline
\multirow{2}{*}{200716C} & 87-118  &  1.67$\pm$0.03   & 20.84$\pm$1.14  & 4.10$\pm$0.11  \\
                         & 118-148 &  1.72$\pm$0.04   & 14.06$\pm$0.99  & 2.70$\pm$0.10  \\ \hline
 \end{tabular}
    \caption{Joint XRT and BAT spectral analysis results for the time bins of GRBs best fit by simple Power Law (PL) model. Displayed parameters include the photon index (\(\Gamma_{XB}\)) and the flux calculated in the 0.3--150 keV energy range for respective time bins. For 160325A (marked with "*"), we found the presence of a strong absorption below 1\,keV. Hence, we report all the spectral parameters in the energy band between 1-10\,keV.}
    \label{tab:xrt_bat_PL}
\end{table}


\begin{table*}[ht]
    \centering
    \renewcommand{\arraystretch}{1.4}
    \setlength{\tabcolsep}{10pt} 
    \begin{tabular}{|c|c|c|c|c|c|c|} \hline
\multirow{4}{*}{GRB}    & \multirow{4}{*}{T - T$_{0}^{\rm BAT}$ \ }& \multicolumn{5}{c|}{Joint fit (XRT+BAT; 0.3-150\,keV)}\\ \cline{3-7}
                        &                                          & \multirow{3}{*}{-$\alpha$}& \multirow{3}{*}{-$\beta$} & \multirow{2}{*}{E$_{\rm p}$} & F$_{0.3-150\,keV}$ & F$_{20\,keV}$\\ 
                        &                                          &                           &               &                 & $\times10^{-9}$    & $\times10^{-9}$           \\
                        &    [s]                                   &                           &               &   (keV)         & [erg\,cm$^{-2}$\,s$^{-1}$] & [erg\,cm$^{-2}$\,s$^{-1}$] \\  \hline
140102A                  & 63-885                                  & 1.46$\pm$0.12             & 2.39$\pm$0.23 & 7.54$\pm$1.52 & 1.86$\pm$0.19    & 0.35$\pm$0.02   \\ \hline
\multirow{1}{*}{150314A} & \multirow{1}{*}{92-778}                 & 1.67$\pm$0.04             & 2.86$\pm$0.21 & 5.87$\pm$0.64 & 2.89$\pm$0.17    &  0.56$\pm$0.02   \\ \hline
\multirow{2}{*}{180720B} & 109-147                                 & 1.14$\pm$0.41             & 2.04$\pm$0.05 & 8.23$\pm$2.23 & 103.00$\pm$3.00  & 16.0$\pm$3.0   \\
                         & 147-625                                 & 1.48$\pm$0.08             & 2.20$\pm$0.06 & 7.84$\pm$0.83 & 15.20$\pm$0.30  &   2.9$\pm$0.05  \\ \hline       
\multirow{1}{*}{210619B} & 334-1864                                & 1.88$\pm$0.03             & 2.73$\pm$0.44 & 9.12$\pm$3.42 & 3.59$\pm$0.29  &  0.64$\pm$0.03 \\   \hline       
\multirow{1}{*}{240825A} & 417-1294                                & 1.54$\pm$0.06             & 2.61$\pm$0.14 & 6.17$\pm$0.52 & 4.53$\pm$0.23  & 0.87$\pm$0.03 \\   \hline
 \end{tabular}
    \caption{Joint XRT and BAT spectral analysis results for the time bins of GRBs best fit by the Smoothly Broken Power Law (SBPL) model. Displayed parameters include the low-energy photon index ($\alpha$), high-energy photon index ($\beta$), and peak energy (E$_{\mathrm{peak}}$) in keV. The flux is calculated in the 0.3--150 keV and at 20 keV.}

    \label{tab:xrt_bat_SBPL}
\end{table*}

\begin{table*}[ht!]
    \centering
    \label{tab:nh_stats}
    \setlength{\tabcolsep}{7.2pt}
    \begin{tabular}{|c|c|c|c|c|c|c|}
        \hline
        \multirow{3}{*}{GRB} & \multirow{2}{*}{T - T$_{0}^{\rm BAT}$ \ }     & \multicolumn{4}{c|}{Joint XRT+BAT}                      & \multicolumn{1}{c|}{XRT} \\ \cline{3-7}
                             &                               & \multicolumn{2}{c|}{PL} & \multicolumn{2}{c|}{SBPL}     & \multicolumn{1}{c|}{PL}  \\ \cline{3-7}
                             &    [s]                        & Stat (dof) & \(N_H\) [$10^{22}$ cm$^{-2}$]   & Stat (dof) & \(N_H\) [$10^{22}$ cm$^{-2}$]         &  \(N_H\) [$10^{22}$ cm$^{-2}$]                \\  \hline
                             
\multirow{2}{*}{081203A}     & 209–640                       & 643 / 690 & 0.21$\pm$0.06  & 643 / 688  & 0.21$\pm$0.06 & 0.22$\pm$0.06 \\
                             & 650-798                       & 142 / 369 & 0.41$\pm$0.38  & 141 / 367  & 0.37$\pm$0.40 & 0.53$\pm$0.42 \\ \hline
                             
        090510               & 100–262                       & 359 / 390 & 0.15$\pm$0.06  & 358 / 388  & 0.16$\pm$0.06 & 0.15$\pm$0.06 \\ \hline
        
        110625A              & 146–240                       & 482 / 525 & 4.43$\pm$0.46  & 449 / 523  & 4.46$\pm$0.49 & 5.04$\pm$0.74 \\ \hline
        
        110731A              & 73–640                        & 778 / 789 & 0.16$\pm$0.12  & 764 / 787  & 0.09$\pm$0.13 & 0.14$\pm$0.12 \\ \hline
        
        120729A              & 393–635                       & 207 / 322 & 0.10$\pm$0.15  & 206 / 320  & 0.08$\pm$0.15 & 0.10$\pm$0.15 \\ \hline
        
        121011A              & 104-138                       & 231 / 436 & $<$0.11        & 230 / 434  & $<$0.10       & 0.05$\pm$0.04 \\ \hline
        
\multirow{3}{*}{130427A}     & 198-301                       & 748 / 701 & 0.05$\pm$0.01  & 747 / 699 & 0.05$\pm$0.01 & 0.07$\pm$0.01\\
                             & 301-463                       & 935 / 812 & $<$0.10        & 922 / 810 & $<$0.10 & $<$0.10\\
                             & 463-500                       & 609 / 648 & $<$0.10        & 584 / 646 & $<$0.10 & $<$0.10\\ \hline
                         
        140102A              & 63-885                        & 1176 / 810& 0.04$\pm$0.01  & 1131 / 808 & $<$0.01 & 0.08$\pm$0.01 \\ \hline
       
\multirow{3}{*}{140323A} & 157-214                           & 371 / 596 & 0.37$\pm$0.04 & 369 / 594 & 0.22$\pm$0.32 & 0.38$\pm$0.04\\
                         & 214-283                           & 244 / 347 & 0.26$\pm$0.06 & 243 / 345 & 0.17$\pm$0.56 & 0.27$\pm$0.06\\
                         & 283-697                           & 271 / 327 & 0.40$\pm$0.07 & 269 / 325 & 0.21$\pm$0.32 & 0.41$\pm$0.07\\ \hline
                         
        150314A          & 92-778                            & 1088 / 866& 0.97$\pm$0.03& 982 / 864  & 0.78$\pm$0.05 & 0.87$\pm$0.04\\ \hline
        
        151006A          & 210-587                           & 568 / 611 & 0.25$\pm$0.03 & 567 / 609 & 0.22$\pm$0.04 & 0.23$\pm$0.04\\ \hline
        
\multirow{2}{*}{160325A} & 73-155                            & 504 / 532 & $<$0.02 & 501 / 530 & $<$0.02 & $<$0.02\\
                         & 230-504                           & 491 / 507 & $<$0.02 & 491 / 505 & $<$0.02 & $<$0.02       \\ \hline   
                         
        160905A          & 65-1110                           & 1038 / 906 & 0.07$\pm$0.01 & 967 / 904 & -- & 0.04$\pm$0.01\\ \hline
        
        160917A          & 102-137                           & 250 / 437 & $<$0.23 & 243 / 435 & $<$0.14 & 0.04$\pm$0.08 \\ \hline
       
\multirow{2}{*}{170405A} & 230-600                           & 640 / 689 & $<$0.13& 627 / 687 & $<$0.14 & $<$0.13 \\
                         & 620-880                           & 294 / 414 & $<$0.78& 268 / 412 & $<$0.81 & $<$0.78\\ \hline
                         
        170728B          & 474-697                           & 554 / 592 & 0.85$\pm$0.10& 546 / 590 & 0.75$\pm$0.11 & 0.87$\pm$0.12\\ \hline
        
\multirow{2}{*}{170813A} & 142-160                           & 134 / 390 & $<$0.27       & 132 / 388 & $<$0.21 & $<$0.22\\
                         & 160-340                           & 126 / 356 & 0.21$\pm$0.11 & 123 / 354 & $<$0.29 & 0.19$\pm$0.11\\ \hline
        
        170906A          & 179-286                           & 573 / 680 & 0.09$\pm$0.02  & 551 / 678 & 0.01$\pm$0.08 & 0.08$\pm$0.02  \\ \hline 
        
        171120A          & 104-229                           & 437 / 506 & 2.21$\pm$0.23  & 432 / 504 & 1.83$\pm$0.26 & 2.00$\pm$0.21\\ \hline

\multirow{2}{*}{180720B} & 109-147                           & 603 / 694  & 0.43$\pm$0.03 & 583 / 692 & 0.22$\pm$0.08 & 0.22$\pm$0.08 \\
                         & 147-625                           & 1218 / 907 & 0.30$\pm$0.08 & 998 / 905 & 0.18$\pm$0.02 & 0.18$\pm$0.02\\ \hline

\multirow{2}{*}{200716C} & 87-118                            & 743 / 706  & 0.16$\pm$0.01 & 673 / 704 & 0.05$\pm$0.02 & 0.09$\pm$0.01 \\
                         & 118-148                           & 696 / 652  & 0.14$\pm$0.01 & 614 / 650 & 0.06$\pm$0.02 & 0.07$\pm$0.01 \\ \hline
                         
\multirow{1}{*}{210619B} & 334-1864                          & 976 / 924  & 0.23$\pm$0.03 & 969 / 922 & 0.20$\pm$0.05 & 0.24$\pm$0.03 \\   \hline 

\multirow{1}{*}{220101A} & 100-239                           & 1205 / 833 & $$<$$0.02         & 1215 / 831 & $$<$$0.02 & $$<$$0.05 \\   \hline

\multirow{1}{*}{240825A} & 417-1294                          & 1086 / 904  & 0.88$\pm$0.02 & 904 / 902 & 0.68$\pm$0.03 & 0.77$\pm$0.02 \\   \hline
    \end{tabular}
    \caption{Spectral fitting statistics and intrinsic absorption (\(N_H\)) values obtained from power-law (PL) and smoothly broken power-law (SBPL) models applied to joint XRT + BAT data and XRT data alone. The table includes the GRB name, time intervals since the BAT trigger (\(T - T_0^{\rm BAT}\)), fit statistics with degrees of freedom (Stat (dof)), and intrinsic absorption column density \(N_H\) for each model.}

    \label{tab:Pl_sbpl}
\end{table*}

\begin{figure}[t]
    \centering
     \includegraphics[width=\columnwidth]{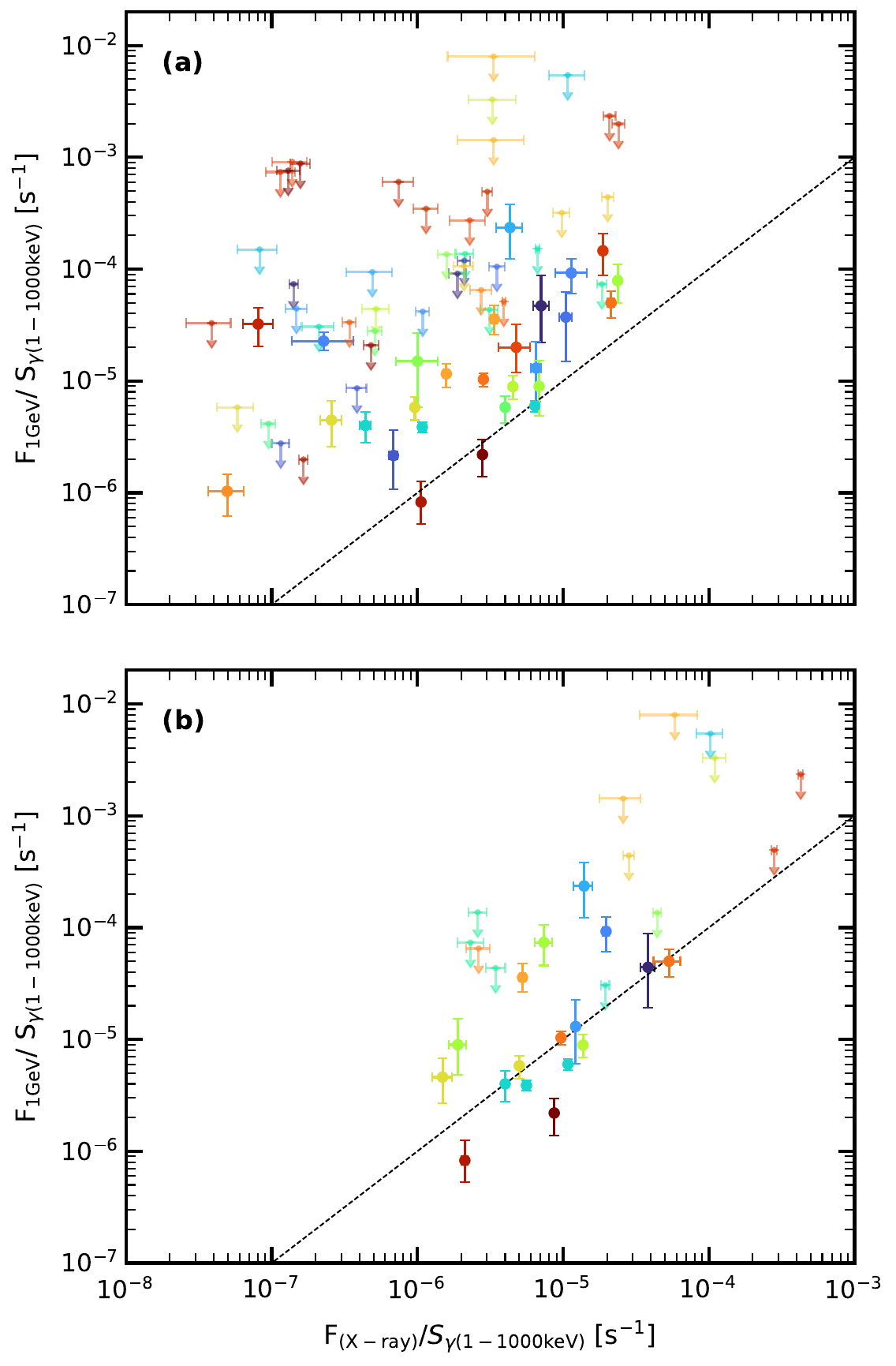}
     \caption{Comparison between X-ray and HE gamma-ray fluxes normalized with fluence (1-1000~keV). Plot \textbf{(a)}: Ratio between flux and fluence in 1 keV  vs. 1 GeV. Plot \textbf{(b)}: Ratio between flux and fluence in 20 keV vs. 1 GeV. The dashed line indicates equality line. Both panels show LAT detections (data points with flux uncertainty, \(\text{TS} > 20\)) and upper limits (downward-pointing with reduced opacity). The right panel has fewer bins than left panel due to limited sensitivity of BAT (see Tab.~\ref{tab:timeresolved} and Sect.~\ref{sec:flux_flux} for details).}
    \label{fig:flux_normalized_XRT_LAT}
\end{figure}

\begin{figure}[H]
    \centering
        \includegraphics[width=\columnwidth, height=5cm]{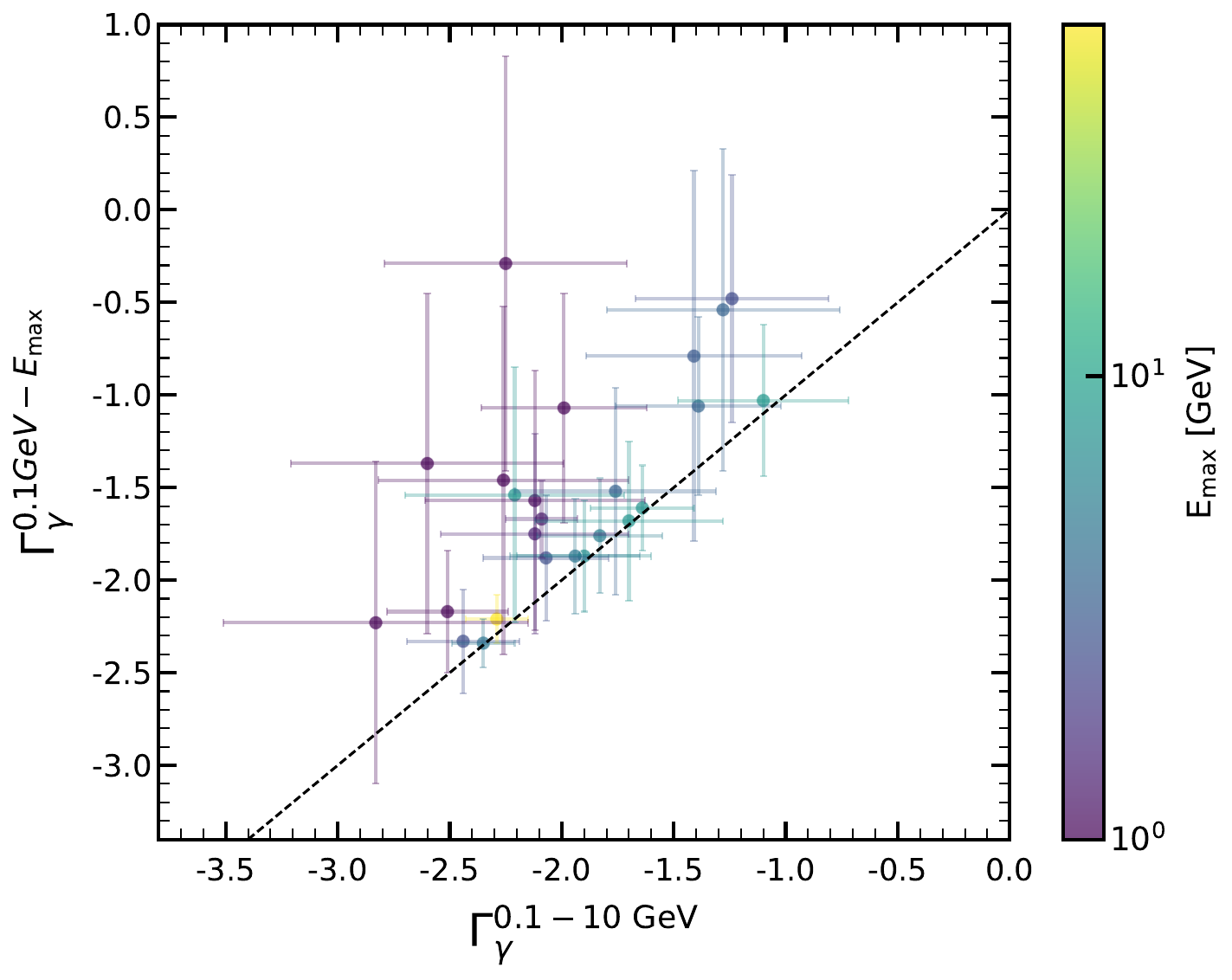}
        \caption{Comparison of spectral indices between HE gamma-rays in the fixed energy range (0.1–10 GeV, x-axis) and the dynamic energy range (0.1\,GeV–E\textsubscript{max}, y-axis). The color map represents the highest photon energy E$\textsubscript{max}$ detected in GeV.}

    \label{fig:Index_LAT_comparison}
\end{figure}

\begin{figure}[H]
    \centering
        \includegraphics[width=\columnwidth, height=6cm]{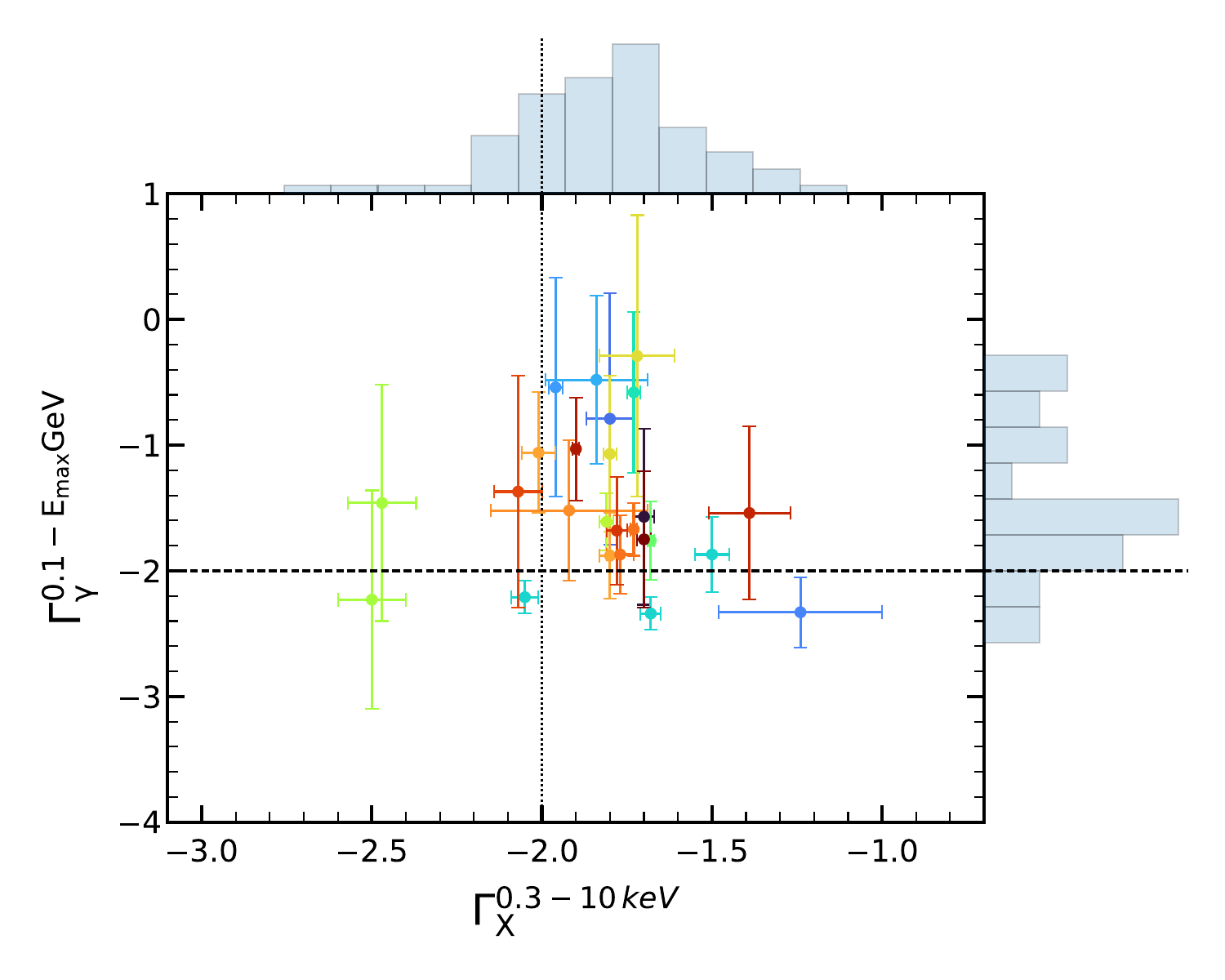}
       \caption{Comparison of spectral indices between X-ray range (0.3–10 keV, x-axis) and the dynamic energy range (0.1\,GeV–E\textsubscript{max}, y-axis). The histograms shown
on the top and right panels display the distributions of photon
indices in the XRT and LAT bands, respectively.}
     \label{fig:Index_EmaxLAT_XRTcomparison}
\end{figure}

\begin{figure}[H]
    \centering
    \includegraphics[width=\columnwidth, height=6cm]{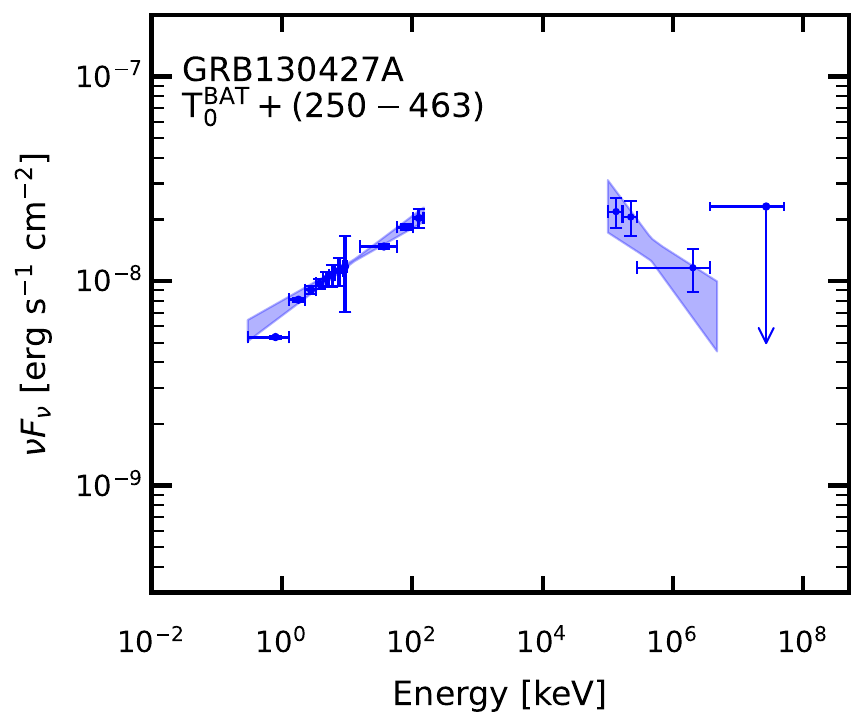}
    \caption{Time-resolved broadband spectrum of GRB130427A for a selected time interval. Butterfly spectra shows the joint spectral fits combining unabsorbed XRT and BAT data in the 0.3--150~keV range, and LAT data from 0.1~GeV up to the highest energy photon (E\(_{\text{max}}\)) within the respective time windows. Spectrum for given bin shows hard X-ray and  soft GeV emission behavior.}
    \label{fig:130427A_SED}
\end{figure}

\begin{figure}[H]
    \centering
     \includegraphics[width=\columnwidth, height=6cm]{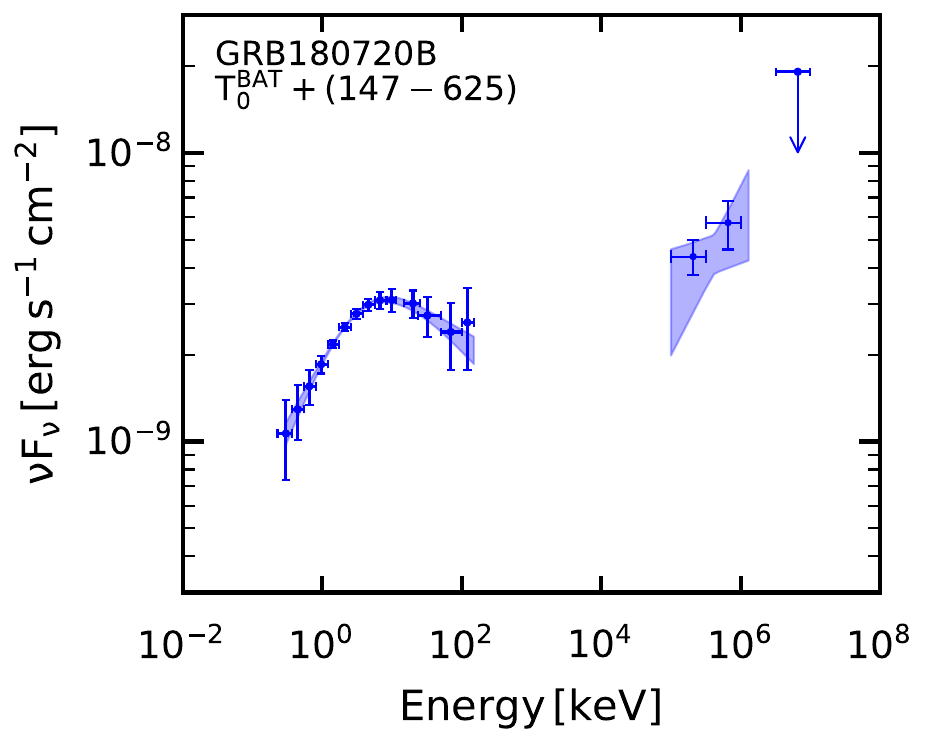}       
     \caption{Time-resolved broadband spectrum of GRB180720B for a selected time intervals. Butterfly shows the joint spectral fits combining unabsorbed XRT and BAT data in the 0.3--150~keV range, and LAT data from 0.1~GeV up to the highest energy photon (E\(_{\text{max}}\)) within the respective time windows. The spectrum illustrate harder X-ray and GeV emission in afterglow phase.}
    \label{fig:180720B_SED}
\end{figure}

\begin{figure}[t]
    \centering
     \includegraphics[width=\columnwidth, height=5.5cm]{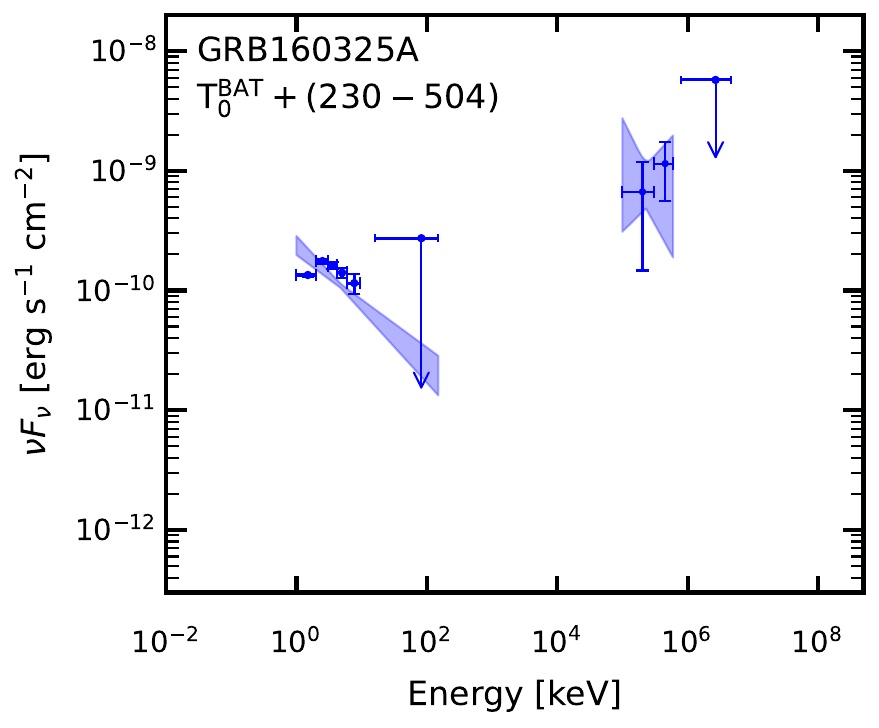}
    \caption{Time-resolved broadband spectrum of GRB160325A for a selected time intervals. Butterfly shows the spectral fits for XRT (1--10~keV) and BAT upper limit in the  15--150~keV range, and LAT data from 0.1~GeV up to the highest energy photon (E\(_{\text{max}}\)) within the respective time windows. The spectrum shows softer X-ray emission with lower emission in X-ray compared to HE gamma-rays.}
    \label{fig:160325A_SED}
\end{figure}

\clearpage

\section{Modeling}
\subsection{Simulation of Long GRBs Sample Based on Fermi/GBM Observations}\label{sec:sims}
\begin{figure}[H]
    \centering
     \includegraphics[width=1\columnwidth]{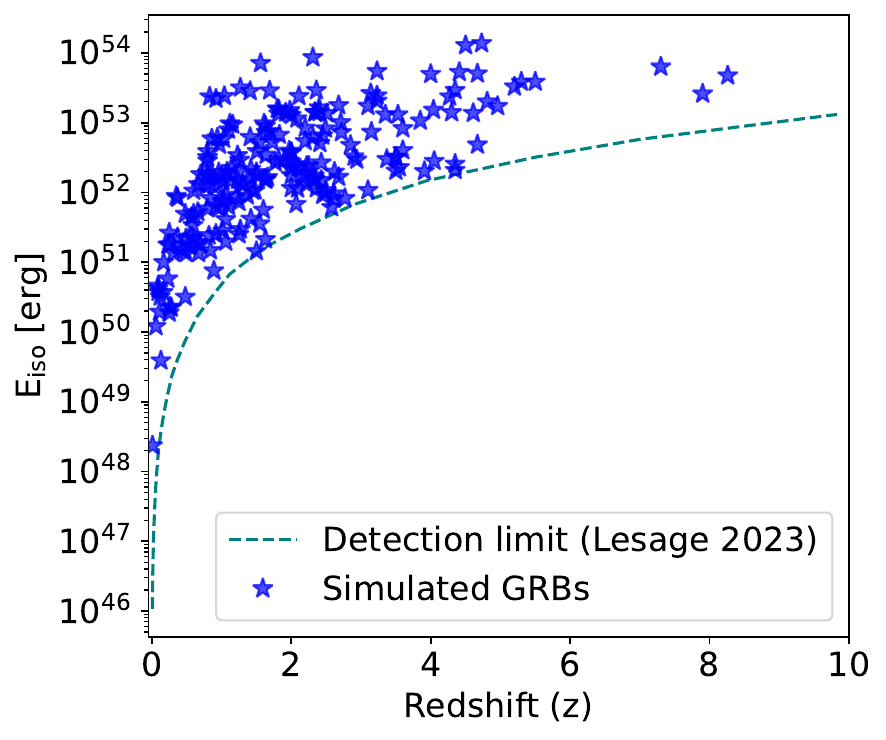}
    \caption{Distribution of a sample of 220 GRBs with randomly assigned \(\mathrm{E}_{\rm ISO}\) and redshift \(z\), based on the distributions reported in \citealt{2022arXiv220606390G}. The dashed line represents the detectability limit of Fermi/GBM, adopted from \citealt{2023ApJ...952L..42L}.}
    \label{fig:Eiso_z}
\end{figure}

We simulated a sample of 220 long GRBs, representing the number of long GRBs typically detected by Fermi/GBM over the course of one year. Figure~\ref{fig:Eiso_z} shows the distribution of the simulated GRBs in the \(E_{\rm ISO}\)--\(z\) plane, where isotropic-equivalent energy \(E_{\rm ISO}\) and redshift \(z\) were randomly assigned following the distributions reported in \citealt{2022arXiv220606390G}. Only GRBs above the Fermi/GBM detectability threshold, as defined in \citealt{2023ApJ...952L..42L}, are included in the simulation.

\subsection{Kolmogorov-Smirnov Test}
To check the comparison between preferred parameters that can explain simultaneously X-ray and GeV photons, we performed the Kolmogorov-Smirnov (KS) test. 
In this work, we performed two independent KS tests. The first compares the predicted gamma-ray luminosity normalized by $E_{\mathrm{iso}}$ (denoted $L_\gamma/E_{\mathrm{iso}}$) with the empirical relation from~\citealt{2014MNRAS.443.3578N}, based on the correlation as a function of the rest frame time. Individual p-values for each 220 points (representing 220 GRB) are derived based on the agreement between the simulated points that satisfies the clustering with a dispersion of 0.23 \citep{2014MNRAS.443.3578N}. We further compute the average p-value in gamma-ray band ($\bar p_{\gamma}$) defined as $\bar p_{\gamma}$=$\Sigma p_{i}$/N$_{tot}$, where p$_{i}$ are the individual probabilities and N$_{tot}$ is the total number of simulated points (220).
Similarly, the average p-value ($\bar p_{x}$) corresponding to the similarity between the distributions of predicted X-ray luminosity normalized by $E_{\mathrm{iso}}$ ($L_X/E_{\mathrm{iso}}$) with that reported in~\citealt{2012MNRAS.425..506D} (within 1$\sigma$ spread) is also evaluated. 
To assess the general agreement between the prediction of the model and the observational correlations in these two bands, we combined the above mentioned p-values ($\bar p_{x}$ and $\bar p_{\gamma}$) of the two KS tests using Fisher’s method, providing a single statistical measure of the compatibility between the model and the observational trends. The results are reported in Tab~\ref{tab:KStest}.

\subsection{VHE vs. X-ray Flux} \label{app:VHE}
As shown in Fig.~\ref{fig:LeMoC_4}, which compares the predicted intrinsic VHE flux with the X-ray flux for the parameter set 
\(p=2.2\), \(\epsilon_{\rm e} = 0.1\), \(\epsilon_{\rm B} = 10^{-4}\), and wind-medium with lower density (A$_{*}$ = 0.1), we include data points for two VHE GRBs: 190114C and 190829A, taken from \citealp{2019Natur.575..455M} and \citealp{2021Sci...372.1081H}, respectively. In \citealp{2019Natur.575..455M}, the reported flux for GRB 190114C is in the energy range 0.3--1~TeV, while for GRB 190829A, \citealp{2021Sci...372.1081H} reports the flux in 0.2--4~TeV. To convert the reported flux to a common energy band 0.3--1~TeV range, we interpolated
the flux assuming the reported intrinsic photon index ($\Gamma_{\rm Y}$), using the following formula:

\begin{equation}
F_{0.3-1\mathrm{TeV}} = F_{e1-e2~\mathrm{TeV}} \times \left[ \frac{1 - \left(0.3/1\right)^{2+\Gamma_{\rm Y}}}{1 - \left(e1/e2\right)^{2+\Gamma_{\rm Y}}} \right]
\end{equation}

Here, e1 and e2 denote the lower and upper bounds of the energy interval reported in given literature, respectively. We calculate flux at
0.3 TeV for comparison at specific energies. This value
is derived from the PL fit performed over the 0.3–1 TeV energy range.

\begin{table*}[ht]
    \centering
    \renewcommand{\arraystretch}{1.5}
    \setlength{\tabcolsep}{16.5pt} 
    \begin{tabular}{|c|c|c|c|c|c|c|} \hline
\multicolumn{4}{|c|}{Preferred parameters}       & \multicolumn{3}{c|}{KS test: p-values}\\ \hline 
     p & $\epsilon_{\rm B}$ &  medium &  density (n/A$_{*}$) & $\bar p_{\gamma}$ &  $\bar p_{x}$                       &{Combined} \\  \hline
\(2.2\) &\(10^{-4}\) & wind        & 0.1       & 0.84& 0.94 & 0.97 \\
\(2.3\)& \(10^{-4}\) & wind        & 0.1       & 0.74& 0.91 & 0.94  \\
\(2.4\)& \(10^{-4}\) & wind        & 0.1       & 0.62& 0.89 & 0.88  \\
\(2.2\) &\(10^{-3}\) & wind        & 0.1       & 0.37& 0.83 & 0.67 \\
\(2.4\) &\(10^{-3}\) & wind        & 0.1       & 0.28& 0.93 & 0.61 \\
\(2.3\) &\(10^{-3}\) & wind        & 0.1       & 0.24& 0.88 & 0.54 \\
\(2.2\)& \(10^{-2}\) & wind        & 1.0       & 0.18& 0.29 & 0.21  \\
\(2.3\)& \(10^{-2}\) & wind        & 1.0       & 0.13& 0.32 & 0.16  \\
\(2.2\) &\(10^{-2}\) & homogeneous & 1.0       & 0.03& 0.12 & 0.02  \\
\(2.3\)& \(10^{-2}\) & homogeneous & 1.0       & 0.02& 0.18 & 0.02 \\
\hline
\end{tabular}
\caption{Kolmogorov–Smirnov (KS) test results for the preferred parameters in both wind and homogeneous (ISM) environments, assuming \(\epsilon_{\rm e} = 0.1\) and $\eta=0.1$.
$\bar p_{x}$ and $\bar p_{\gamma}$ represent the probability of agreement between 220  simulated GRBs and the clustering in X-rays and GeV energies, respectively. The last column report the joint probability.}

\label{tab:KStest}
\end{table*}

\begin{figure*}[ht!]
    \centering
     \includegraphics[width=1\columnwidth]{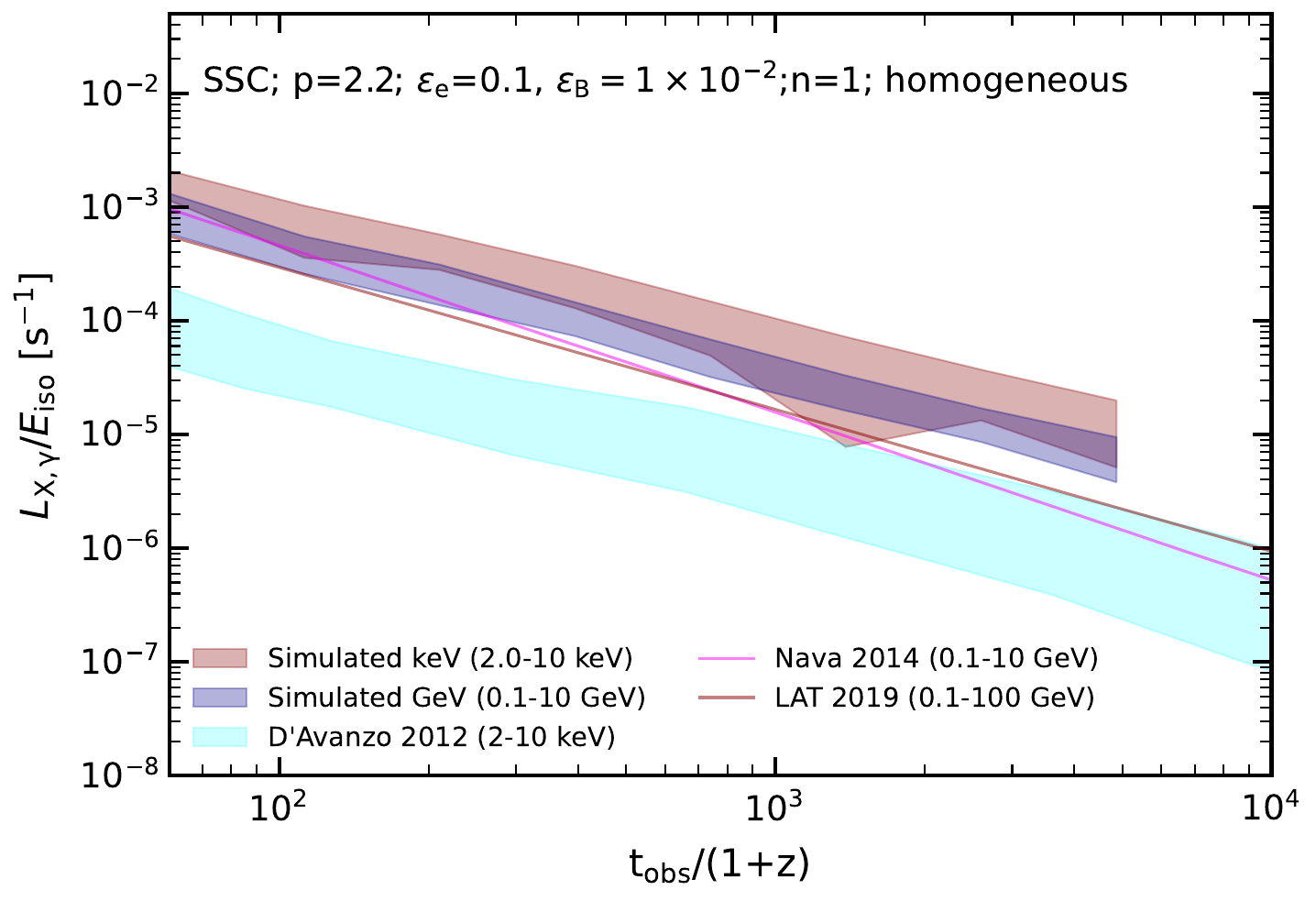}
     \includegraphics[width=1\columnwidth]{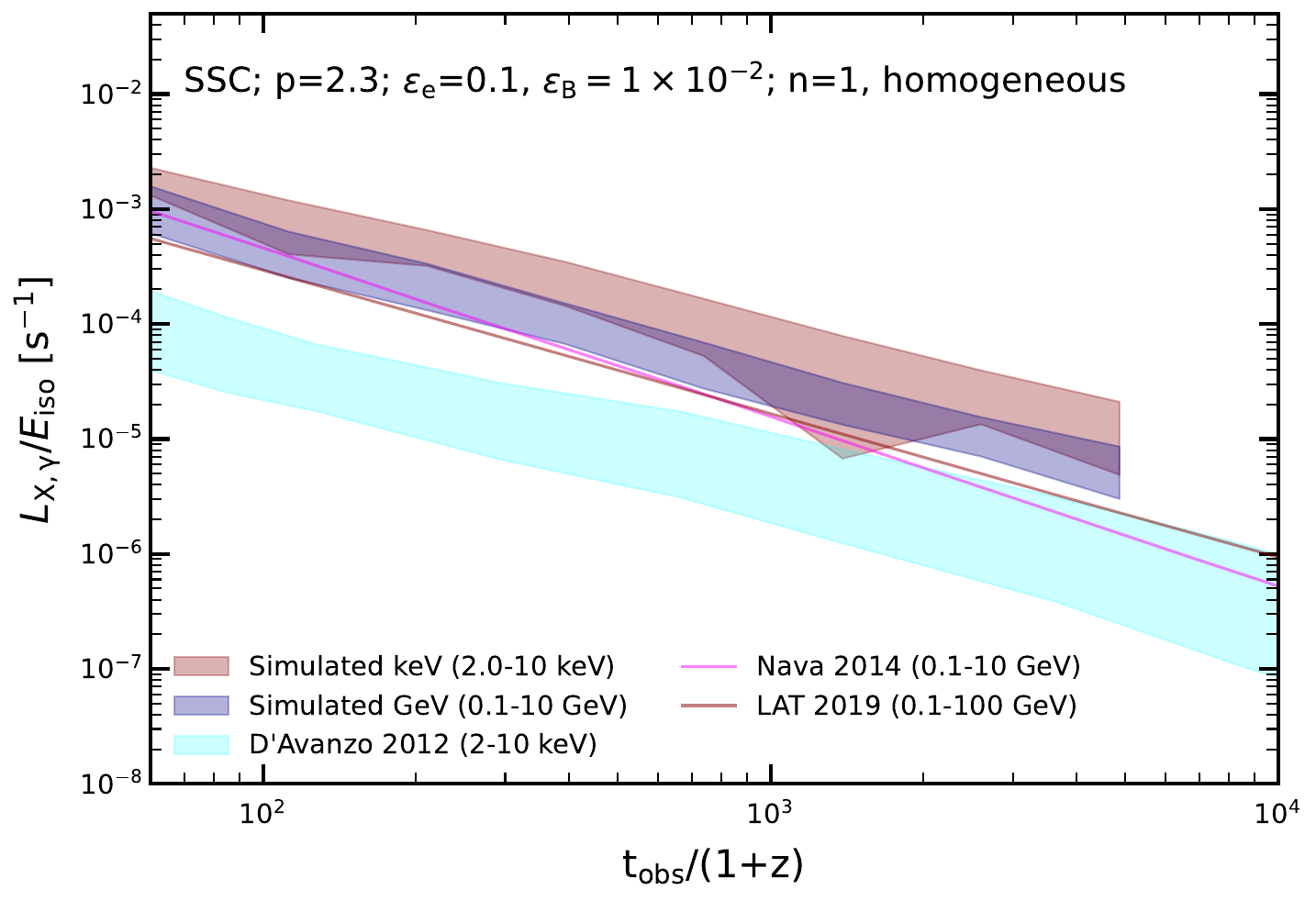}
     \includegraphics[width=1\columnwidth]{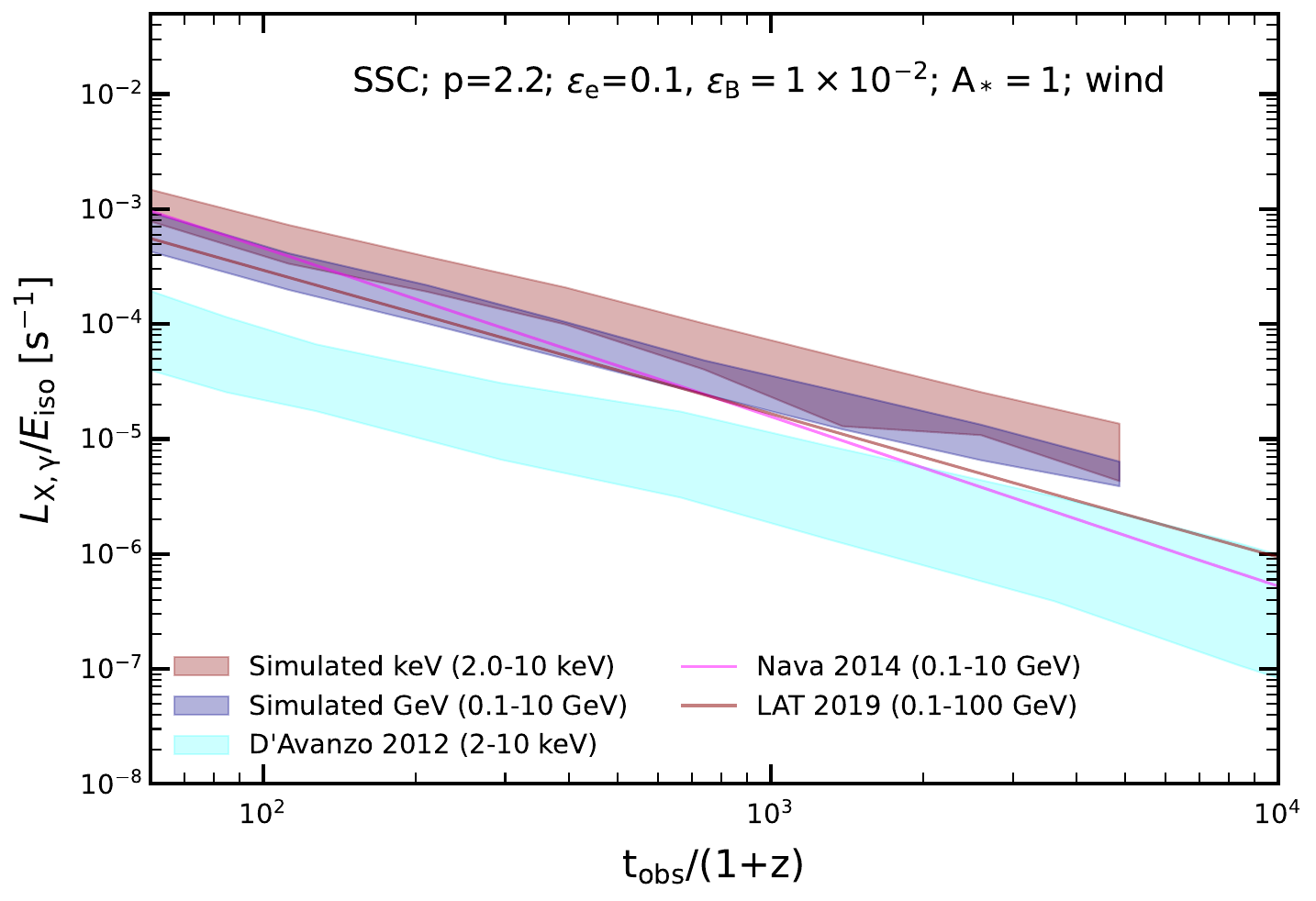}
     \includegraphics[width=1\columnwidth]{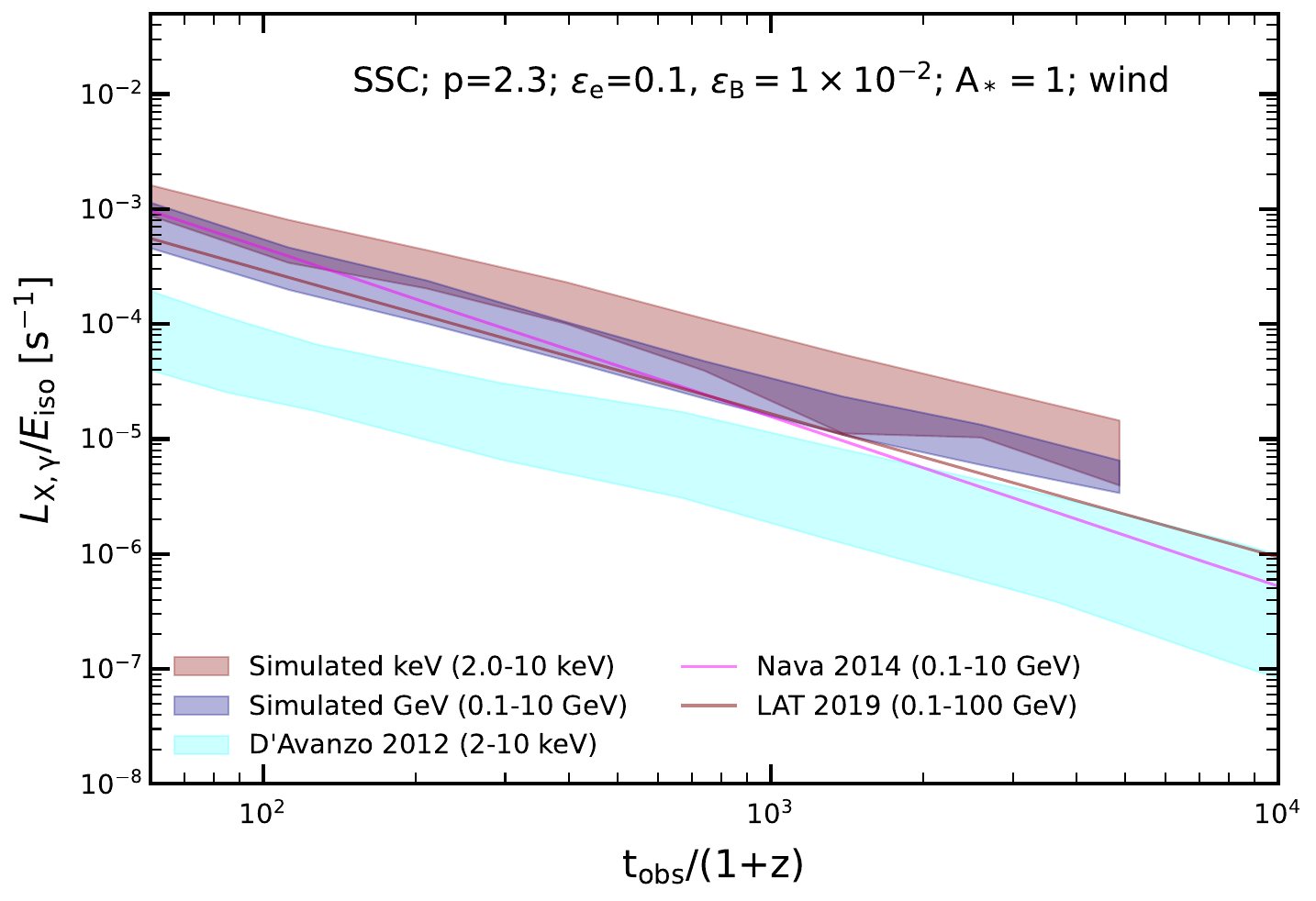}

     \caption{The relation between the luminosity in X-ray ($L_{\rm X}$)  and HE gamma-rays ($L_{\gamma}$) with the prompt emission isotropic energy (E$_{\rm iso}$) in rest frame of GRBs. The solid pink and magenta lines represent the trends reported by~\citealt{2014MNRAS.443.3578N} and \citealt{2019ApJ...878...52A}, respectively, for the ratio L$_{\gamma}/E_{\rm iso}$. The shaded cyan band indicates the $1\sigma$ confidence region for the corresponding trend in X-rays (L$_{\rm X}/E_{\rm iso}$) reported in \citealt{2012MNRAS.425..506D}, where $L_{\rm X}$ is calculated in the 2--10\,keV band. The magenta and blue band represents the simulated GeV and X-ray emission for different set of microphysical parameters explained in Sect.\ref{sec:preferred_model}.}
\label{fig:LaraPaolo}    
\end{figure*}

\end{document}